\documentclass[a4paper,11pt]{article}
\pdfoutput=1 
\usepackage{jheppub}

\usepackage[T1]{fontenc} 
\usepackage{blindtext}
\usepackage{hyperref}
\usepackage{amsmath,amssymb}
\usepackage{float}
\usepackage{microtype}
\usepackage{graphicx}
\usepackage{bm}
\usepackage{latexsym}
\usepackage{epsfig}
\usepackage{psfrag}
\usepackage{color}
\usepackage[dvipsnames]{xcolor}
\usepackage{subfigure}
\usepackage{graphicx}
\usepackage{tikz}
\usepackage{tikz-feynman}
\tikzfeynmanset{compat=1.1.0}
\usepackage{amssymb}
\usepackage{amsmath}
\usepackage{bm}
\usepackage{latexsym}
\usepackage{epsfig}
\usepackage{psfrag}
\usepackage[normalem]{ulem}
\usepackage{textcomp}
\usepackage{color}
\usepackage{pstricks}
\usepackage[utf8]{inputenc}
\usepackage{tcolorbox}
\usepackage{feynmp-auto}
\usepackage{tikz-feynman}
\usepackage{simpler-wick}
\usepackage{outlines}
\usepackage{physics}
\usepackage{comment}
\usetikzlibrary{decorations.pathmorphing,decorations.markings}
\usetikzlibrary{positioning}
\usetikzlibrary{arrows,positioning,decorations.markings}\tikzset{->-/.style={decoration={markings,mark=at position #1 with {\arrow{>}}},postaction={decorate}}}

\newenvironment{equations}{\equation\aligned}{\endaligned\endequation}  
\newcommand{\beq}{\begin{equations}}
\newcommand{\eeq}{\end{equations}}
\newcommand{\bc}{\begin{cases}}
\newcommand{\ec}{\end{cases}}
\newcommand{\boe}{\begin{outline}[enumerate]}
\newcommand{\eo}{\end{outline}}

\def\l{\left(}
\def\r{\right)}
\def\ll{\left[}
\def\rr{\right]}
\def\nn{\nonumber}

\def\={&=}
\def\p{\partial}

\def\-{\item[-]}

\def\wc{W^{\zeta}}

\newcommand{\bk}{\boldsymbol{k}}

\makeatletter
\gdef\@fpheader{}
\makeatother

\title{Inflationary trispectrum of gauge fields from scalar and tensor exchanges}
\author{P. Jishnu Sai,}
\author{S. R. Haridev}
\author{and Rajeev Kumar Jain}
\affiliation{Department of Physics, Indian Institute of Science\\
C. V. Raman Road, Bangalore 560012, India}
\emailAdd{jishnup@iisc.ac.in}
\emailAdd{haridevsr@iisc.ac.in}
\emailAdd{rkjain@iisc.ac.in}

\abstract{
In this paper, we compute the inflationary trispectrum of primordial gauge fields generated through the scalar and tensor exchanges in models with spectator $U(1)$ gauge fields which are kinetically coupled to the inflaton.
Focusing on the connected four-point autocorrelation function of gauge fields, we derive exact analytical expressions for the full trispectrum of both electric and magnetic fields using the in-in formalism and cosmological diagrammatic rules, and explore their respective contributions in specific momentum configurations. For the scalar exchange, we find that the trispectrum signal in the equisided configuration grows with the exchange momentum and reaches its maximum in the flattened limit.
However, in the counter collinear limit, we show that the non-linearity parameter associated with the trispectrum scales quadratically with the corresponding parameter of the cross-correlation bispectrum of magnetic fields and curvature perturbations, thereby establishing a hierarchical relation between the higher- and lower-order correlation functions.
For the tensor exchange, the trispectrum displays a richer angular dependence, reflecting the sensitivity to the orientation of the momentum quadrilateral with respect to the tensor polarisation, producing characteristic angular modulations in the trispectrum. 
Detecting such angular signatures in future high-precision cosmological observations would provide a novel window into tensor-mediated interactions in the early universe.
}

\begin{document} 
\maketitle
\flushbottom
\section{Introduction}

Precision cosmological observations provide us with a remarkable opportunity to probe the fundamental laws of physics at extremely high energies. The inflationary epoch of the early universe, operating at energy scales only a few orders of magnitude below the Planck scale, provides a natural setting to explore the perturbative regime of quantum gravity \cite{Starobinsky:1980te, Guth:1980zm, Linde:1981mu, Starobinsky:1979ty, Mukhanov:1981xt, Sriramkumar:2009kg, Martin:2013tda}. Inflationary dynamics leave telltale imprints on primordial correlation functions involving scalar and tensor fluctuations, which can be strongly constrained by cosmological observations \cite{Martin:2013nzq, Planck:2018jri, Chowdhury:2019otk}. 
Although the two-point correlation function, or power spectrum, of scalar perturbations is tightly constrained by observations, primordial tensor perturbations generated during inflation have not yet been measured, with current data providing only upper limits on their amplitude~\cite{BICEP:2021xfz, Campeti:2022vom}. One of the primary goals of the forthcoming cosmological surveys is to detect tensor perturbations as well as higher-point scalar correlators and correlators involving tensor perturbations \cite{Gangui:1999vg, Wang:1999vf, Maldacena:2002vr, Creminelli:2003iq, Lyth:2005fi, Seery:2005wm, Chen:2006nt, Chen:2010xka, Wang:2013zva, Kristiano:2021urj}. 
Since a wide range of inflationary models yield observationally indistinguishable predictions at the level of the power spectrum, precision measurements of higher-order correlation functions become essential for breaking this degeneracy and provide us with decisive clues to underpin the physics of inflation and the initial conditions of the very early universe
\cite{Planck:2019kim, LiteBIRD:2022cnt, Achucarro:2022qrl}.

The primordial trispectrum of comoving curvature perturbations $\zeta$ during inflation encodes crucial information about the dynamics and interactions of the inflationary epoch beyond what is accessible through the power spectrum or bispectrum alone. The amplitude and shape of the trispectrum provide novel insights into the underlying interacting physics during inflation, sourcing non-Gaussianities due to e.g., higher-derivative interactions,  exchange of light scalar or tensor fields, or the presence of multiple fields with non-trivial interactions \cite{Seery:2006vu, Chen:2006dfn, Arroja:2008ga, Seery:2008ax, Chen:2009bc, Elliston:2012wm}.   
For example, in single-field slow-roll inflationary models, the trispectrum is typically small and local in shape, which is suppressed by slow-roll parameters~\cite{Seery:2006vu}. In contrast, a trispectrum with distinctive shapes, such as equilateral or collapsed configurations, and potentially observable amplitudes
will hint toward beyond single-field slow-roll models. In particular, exchange diagrams involving light scalars or tensors can also imprint non-trivial momentum dependence, making the trispectrum a sensitive probe of hidden sectors and particle content during inflation~\cite{Byrnes:2013qjy, Planck:2019kim,
Philcox:2025bvj, Philcox:2025lrr, Philcox:2025wts}. The CMB trispectrum induced by primordial magnetic fields, and the resulting constraints on the magnetic field strength, have been discussed in \cite{Trivedi:2011vt, Trivedi:2013wqa}. 

Beyond single-field inflationary models, non-trivial shapes of primordial correlation functions can serve as a valuable probe of the presence of multiple dynamical fields during inflation \cite{Bernardeau:2002jy, Rigopoulos:2005ae}.
An interesting multi-field scenario involves primordial gauge fields, which can become dynamical during inflation if their conformal symmetry is broken. Moreover, if these gauge fields are identified with the canonical electromagnetic fields, such a scenario would also lead to the generation of large scale primordial magnetic fields in the universe \cite{Turner:1987bw, Ratra:1991bn, Subramanian:2009fu, Durrer:2013pga}. One of the widely explored scenarios is the kinetic coupling model, in which the kinetic term of the gauge fields is directly coupled to the inflaton through a dilatonic coupling \cite{Bamba:2006ga, Martin:2007ue, Demozzi:2009fu, Ferreira:2013sqa, Ferreira:2014hma}. Note that, such a coupling only breaks the conformal invariance while preserving the gauge invariance. Such scenarios not only generate cosmological magnetic fields of relevant strength on large scales as required by blazar observations but also induce non-trivial cross-correlations between scalar and tensor perturbations and magnetic fields which turn out to be strongly non-Gaussian in nature. A lot of work has been done in this direction to understand the full bispectrum, higher-order cross-correlations and associated soft theorems \cite{Caldwell:2011ra, Motta:2012rn, Jain:2012ga, Shiraishi:2012xt, Jain:2012vm, Kunze:2013hy,Fujita:2013qxa,Nurmi:2013gpa,Chowdhury:2018blx,Chowdhury:2018mhj,Jain:2021pve,Sai:2023vyf, Tripathy:2024ngu}. 

In this work, we go beyond the cross-correlation bispectrum and compute the trispectrum of primordial gauge fields generated through the scalar and tensor exchanges during inflation. Since the energy density of gauge fields is quadratic in nature, only even-point autocorrelation functions survive -- such as the two-point function, which gives the power spectrum, and the connected four-point function, which gives the trispectrum -- while all odd-point functions vanish identically. Consequently, non-trivial information about the dynamical evolution of gauge fields, governed by the form of the coupling function, is encoded in the trispectrum. Furthermore, as we shall discuss in detail, the signatures of exchange interactions mediated by scalar or tensor perturbations are contained exclusively in the trispectrum. It is therefore both important and natural to compute the trispectrum in these models, as it provides an information-rich correlation function beyond the power spectrum.

To further highlight the relevance of the trispectrum, we first express the energy density $\rho_{B}(\boldsymbol{k})$ of the magnetic field $B_{i}(\boldsymbol{k})$ as
\begin{equation}
  \rho_{B}(\boldsymbol{k}) \;=\; \int\!\frac{d^{3}p}{(2\pi)^{3}}\;
  B_{i}(\boldsymbol{p})\,B_{i}(\boldsymbol{k}-\boldsymbol{p})\,,
\end{equation}
such that
\begin{align}
  \big\langle \rho_{B}(\boldsymbol{k})\,\rho_{B}(\boldsymbol{k}') \big\rangle
  &= \int\!\frac{d^{3}p}{(2\pi)^{3}}\!\int\!\frac{d^{3}p'}{(2\pi)^{3}}\;
     \big\langle B_{i}(\boldsymbol{p})\,B_{i}(\boldsymbol{k}-\boldsymbol{p})\,
    B_{j}(\boldsymbol{p}')\,B_{j}(\boldsymbol{k}'-\boldsymbol{p}')\big\rangle \nonumber\\[2pt]
  &=\, (2\pi)^{3}\delta^{(3)}(\boldsymbol{k}+\boldsymbol{k}')\,
     \Big[\,P^{\rm disconn}_{\rho_B}(k)+P^{\rm conn}_{\rho_B}(k)\,\Big],
\end{align}
with \(k\equiv|\boldsymbol{k}|\).
The two distinct contributions identified above correspond to the disconnected and connected parts of the trispectrum, which can be expressed as

\medskip
\noindent\textbf{Disconnected or Gaussian term:}
By applying Wick's theorem and using the definition
$
\langle B_{i}(\boldsymbol{k}) B_{j}(\boldsymbol{k}')\rangle
=(2\pi)^{3}\delta^{(3)}(\boldsymbol{k}+\boldsymbol{k}')\,\Pi_{ij}(\hat{\boldsymbol{k}})\,P_{B}(k)
$,
then the disconnected contribution for non-zero $k$ is, 
\begin{align}
  P^{\rm disconn}_{\rho_B}(k)
  &= 2 \int\!\frac{d^{3}p}{(2\pi)^{3}}\;
     \Big[\Pi_{ij}(\hat{\boldsymbol{p}})\,\Pi_{ij}\!\big(\widehat{\boldsymbol{k}-\boldsymbol{p}}\big)\Big]\,
     P_{B}(p)\,P_{B}(|\boldsymbol{k}-\boldsymbol{p}|)\,.
\end{align}
This can be fully computed using the power spectrum $P_B$ and establishes the Gaussian baseline, while all non-Gaussian dynamics enter through \(P^{\rm conn}_{\rho_B}\).

\medskip
\noindent\textbf{Connected or exchange term:}
Expressing the magnetic trispectrum as
\begin{align}
\langle B(\boldsymbol{k}_{1})\!\cdot\!B(\boldsymbol{k}_{2})\,
B(\boldsymbol{k}_{3})\!\cdot\!B(\boldsymbol{k}_{4})\rangle
= (2\pi)^3 \delta^{(3)}(\bk_1+\bk_2+\bk_3+\bk_4)\mathcal{T}_{B}(\boldsymbol{k}_{1},\boldsymbol{k}_{2},\boldsymbol{k}_{3},\boldsymbol{k}_{4})~,
\end{align}
we can write,
\begin{align}
  P^{\rm conn}_{\rho_B}(k)
  &= \int\!\frac{d^{3}p}{(2\pi)^{3}}\!\int\!\frac{d^{3}p'}{(2\pi)^{3}}\;
\mathcal{T}_{B} \big(\boldsymbol{p},\,\boldsymbol{k}-\boldsymbol{p},\,\boldsymbol{p}',\, -\boldsymbol{k}-\boldsymbol{p}'\big).
\end{align}
This contribution arises solely from the exchange interactions. Schematically, one can write the total exchange trispectrum from scalar and tensor exchange as
\begin{align*}
\mathcal{T}_{B} =\mathcal{T}_{\zeta B}+\mathcal{T}_{\gamma B}~.
\end{align*}
Each of these contributions, when evaluated using the in-in formalism, yields a combination of nested time-integrals with corresponding coefficients. In the subsequent sections, we discuss the intricacies of these integrals and the non-trivial structure of coefficients that arise from the polarisation of gauge fields. 
Since the pivotal information about exchange processes is only contained in the connected part, the trispectrum emerges as a unique probe of such interactions, motivating us to compute the inflationary trispectrum of gauge fields arising from scalar and tensor exchange interactions. To our knowledge, our work provides the first complete calculation of these contributions, thereby opening a new avenue to probe the gauge field dynamics during inflation.

This paper is organized as follows: In the next section, we briefly discuss the dynamics of primordial scalar and tensor perturbations and the evolution of gauge fields during inflation. In Section \ref{section:3}, we outline the essentials of the in-in formalism and compute the cubic order interaction Hamiltonians to calculate the higher-order correlators. In Section \ref{section:4}, we present our main results of the trispectrum of gauge fields arising from the exchange of scalar and tensor perturbations and explore the equisided and counter collinear configurations of the magnetic field trispectrum. We briefly comment on the electric field trispectrum as it remains subdominant for the growing kinetic coupling. Finally, in Section \ref{conclusions}, we summarise our results and conclude with a future outlook. The appendices provide supporting details: Appendix \ref{CDrules} outlines the cosmological diagrammatic rules, Appendix \ref{Polcoeffs} lists the polarisation coefficients of the magnetic trispectra in different channels, Appendix \ref{intresults} presents the integral results for various values of $n$, and Appendix \ref{intidentity} contains the evaluation of an integral identity using the equations of motion of the gauge fields.

Throughout this paper, we work in natural units with $\hbar = c =1$, and the Planck mass $M_{\rm Pl}^2 =1/8\pi G$ is set to unity. Our metric convention is $(-,+,+,+)$.

\section{Inflationary dynamics of primordial perturbations and gauge fields}
\label{eq:mps}

In this section, we present a brief overview of the dynamics of scalar and tensor perturbations, along with the evolution of electromagnetic fields during inflation.
We begin by reviewing inflationary scalar perturbations that are well described in terms of the comoving curvature perturbation $\zeta$, which is determined by the Mukhanov-Sasaki equation \cite{Mukhanov:1985rz, Sasaki:1986hm} 
\beq 
\label{eom_R}
\zeta''+2 \frac{z'}{z}\zeta'-\nabla^2\zeta=0~,
\eeq 
where $z=a\sqrt{2\,\epsilon}$, $a$ is the scale factor, $\epsilon$ is the first slow-roll parameter, and $\nabla^2$ denotes the spatial Laplacian in comoving coordinates. The standard quantisation procedure through mode expansion leads to the following solution of mode function of the comoving curvature perturbation,  
\begin{align}
\zeta_k(\eta) = \frac{1}{\sqrt{2\epsilon}}\frac{H}{\sqrt{2k^3}}\left(1+i k \eta\right){\rm e}^{-ik\eta} ~.
    \label{eq: scalar mode 1}
\end{align}
Evidently, the two-point correlation function of the curvature perturbation and corresponding power spectrum are determined by this mode function, i.e.,  $P_{\zeta}=|\zeta_k(\eta)|^2$. Similarly, the Bunch-Davies normalized solution for the Fourier mode function $\gamma_{k}(\eta)$ of tensor perturbation is
\beq
\label{modefn_g}
\gamma_k(\eta) = \frac{H}{\sqrt{k^3}}(1+ik\eta)\,e^{-ik\eta} ~,
\eeq
with the dimensionless power spectrum ${\cal P}_{\gamma}$ as
\beq
\label{ps-dimensionless}
{\cal P}_{\gamma} \equiv \frac{k^3}{2 \pi^2} P_{\gamma} (k,\eta) = 2\frac{k^3}{2 \pi^2}|\gamma_k (\eta)|^2,
\eeq 
which, in the super-horizon limit, leads to ${\cal P}_{\gamma} \sim (H/2 \pi)^2$. The tensor power spectrum thus measures the energy scale during inflation and is exactly scale invariant in a de-Sitter background, and nearly scale invariant in slow-roll inflation.  

In the context of inflationary magnetogenesis, one of the widely studied model is the kinetic coupling model, which breaks the conformal invariance while maintaining gauge invariance by coupling the gauge fields directly to the inflaton. The modified EM action is thus given by  
\beq
 S[A_{\mu}] =- \frac{1}{4}\int {\rm d}^4x\sqrt{-g}~\lambda(\phi)F_{\mu\nu}F^{\mu\nu},
 \label{eq: EM action}
\eeq
where $F_{\mu\nu}=\partial_{\mu}A_{\nu}-\partial_{\nu}A_{\mu}$ is the field strength tensor of the gauge field $A_{\mu}$, and
$\lambda(\phi)$ is the direct coupling between the gauge field and the inflaton, which is often referred to as the kinetic coupling or the dilatonic coupling. 
Various forms of this coupling function have been studied in the context of inflationary magnetogenesis
 \cite{Turner:1987bw,  Ratra:1991bn, Martin:2007ue, Durrer:2010mq, Jain:2012jy, Byrnes:2011aa, Ferreira:2013sqa, Ferreira:2014hma, Tripathy:2021sfb, Tripathy:2022iev}. 
 
Following a similar quantization procedure as in the case of scalar and tensor perturbations, the Fourier mode functions of gauge field fluctuations $A_k$ obey the following equation
\beq
\label{EOM_A}
A^{\prime \prime\sigma}_k+\frac{\lambda'}{\lambda}A^{\prime\sigma}_k+k^2A^\sigma_k=0,
\eeq
where $\sigma$ represents the two physical polarisation states of the gauge field. In the absence of any parity-violating terms in the action, as in the case of the kinetic coupling scenario, the mode functions for both polarisations follow the same equation of motion.  Therefore, for brevity, we will omit the superscript $\sigma$ from $A^\sigma_k$ in the remainder of the paper. It is evident from the above equation that the evolution of gauge fields is completely determined by the choice of the coupling function. 
For instance, cosmological seed magnetic fields from inflation have been studied with an exponential coupling function of the form $\lambda(\phi) \sim e^{2\phi}$ \cite{Ratra:1991bn}. However, as in earlier works, we shall parameterize the time dependence of the coupling function as a power-law in terms of the scale factor as 
\cite{Martin:2007ue, Demozzi:2009fu,Ferreira:2013sqa,Ferreira:2014hma, Tripathy:2021sfb} 
\beq 
\label{coupling}
\lambda\l\phi(a)\r= \lambda_e\left(\frac{a}{a_e}\r^{2n}~,
\eeq 
where $a_e$ and $\lambda_e$ are the scale factor and the coupling at the end of inflation when $\eta = \eta_e$. 
Subsequently, in the slow-roll regime with $a(\eta)\simeq -1/(H\,\eta)$, the gauge field mode function can be obtained by substituting eq. (\ref{coupling}) into eq. (\ref{EOM_A}) and the complete Bunch-Davies normalized solution is given by
\begin{eqnarray}
A_k(\eta) = \frac{1}{\sqrt{\lambda_e}}\frac{\sqrt{\pi}}{2}e^{i\pi(n+1)/2}\sqrt{-\eta}\left(\frac{\eta}{\eta_e}\right)^n H^{(1)}_{n+\frac{1}{2}}(-k\eta) \label{modfn_A}~,
\end{eqnarray}
where $H^{(1)}_{\nu}(x)$ is the Hankel function of the first kind with index $\nu$.
As earlier, one can define the two-point functions of the gauge field, 
\begin{eqnarray} 
 \left\langle A_i(\eta,{\bf k}) A_j(\eta,{\bf k}') \right\rangle= (2\pi)^3 \delta^{(3)}({\bf k}+{\bf k}')\left(\delta_{ij}-\frac{ k_i  k_j}{k^2}\right) \left| A_k(\eta)\right|^2.
 \end{eqnarray} 
The associated magnetic and electric power spectra, denoted as $P_B$ and $P_E$ can be expressed in terms of the gauge field mode function and its time derivative as \cite{Jain:2021pve}
\beq
P_B=2\, \frac{k^2}{a^4} |A_k(\eta)|^2, \qquad P_E=\frac{2}{a^4}|A'_k(\eta)|^2 , 
\eeq
which can further be related to the spectral energy densities of the magnetic and electric fields, often denoted as $\mathcal{P}_B$ and $\mathcal{P}_E$, as 
\begin{eqnarray} 
\mathcal{P}_B \equiv \frac{d \rho_B}{d\ln k}=\frac{\lambda k^3}{4\pi^2}P_B~, \qquad  \mathcal{P}_E \equiv \frac{d \rho_E}{d\ln k}=\frac{\lambda k^3}{4\pi^2}P_E~.
\end{eqnarray} 
The spectral energy densities depend strongly on the value of $n$ in the choice of coupling function. For instance, it is well known that the magnetic spectral energy density becomes scale invariant for both $n=2$ and $n=-3$. However, the former suffers from the so-called strong coupling problem, while the latter is plagued by the strong backreaction problem arising from the electric fields. 
\section{In-in formalism and cubic order interaction Hamiltonians}
\label{section:3}

In all scenarios wherein the gauge fields are directly coupled to the inflaton, such as in the kinetic coupling model in eq. (\ref{eq: EM action}), it is natural to expect cross-correlations between gauge field fluctuations and inflationary scalar and tensor perturbations. At the leading order, these are the three-point cross-correlations involving one of the scalar or tensor perturbations and two powers of the gauge fields which have been studied extensively during inflation in \cite{ Caldwell:2011ra, Motta:2012rn, Jain:2012ga,Shiraishi:2012xt,
Jain:2012vm,Chowdhury:2018blx, Chowdhury:2018mhj,Jain:2021pve}.

In this section, we briefly outline the standard formalism of computing equal-time correlation functions in the early universe, known as the in-in formalism. We shall also present the computations of the interaction Hamiltonian at cubic order with inflationary metric perturbations.

\subsection{The in-in formalism}
\begin{figure}[!t]
\centering 
\begin{tikzpicture}[>=stealth]
  \draw[->, thick] (-6,0) -- (1,0) node[below right] {$\mathrm{Re}(\eta)$};
  \draw[->, thick] (0,-2) -- (0,2) node[above left] {$\mathrm{Im}(\eta)$};

\draw[thick, black] (-6,0.7) -- (-0.9,0)node[pos=0,above=1pt] {$-\infty( 1+i\epsilon)$};
\draw[thick, black] (-6,-0.7) -- (-0.9,0)node[pos=0,below=1pt] {$-\infty( 1-i\epsilon)$};
\node at (-3.5,0.64) {\textcolor{black}{$\mathcal{C}_+$}};
\node at (-3.5,-0.7) {\textcolor{black}{$\mathcal{C}_-$}};

\filldraw[black!60!black] (-0.95,0) circle (2pt);
\node[right, black!50!black] at (-1.3,-0.5) {$\mathcal{O}(\eta_0)$};
\end{tikzpicture}
\caption{Time contour in the in-in formalism, prescribing the initial Bunch–Davies vacuum state.}
\label{tcontour}
\end{figure}
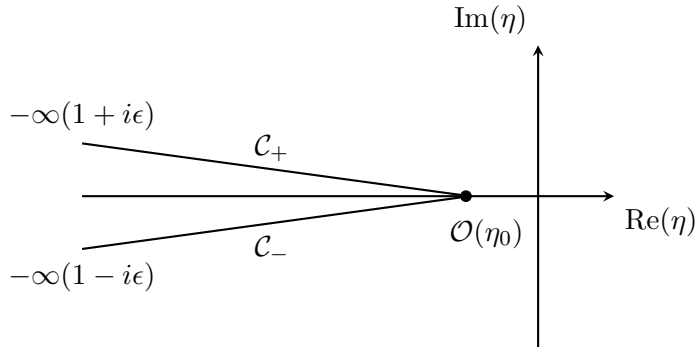

The equal-time quantum correlations in the early universe are computed using the in-in formalism. In this framework, one systematically computes the expectation value of a quantum operator $\mathcal{O}$ at time $\eta_0$, using the in-in master formula, as
\begin{equation}
    \label{exp1_in-in}
    \left\langle \mathcal{O}(\eta_0)  \right\rangle = \left\langle
    0\left|
    {\bar T}\left(e^{i\int_{-\infty}^{\eta_0}d\eta\, H_{\rm int}}\right) \mathcal{O}(\eta_0) T \left(e^{-i\int_{-\infty}^{\eta_0} d\eta\, H_{\rm int}}\right)\right|
    0 \right\rangle,
\end{equation}
where $H_{\rm int}$ is the interaction Hamiltonian and $T$, $\bar T$ are the time ordering and anti-time ordering operators, respectively. The quantum state $\left| 0 \right\rangle$ refers to the initial vacuum state of the theory, which we take to be the standard Bunch-Davies vacuum state. Here, the “in” vacuum state is selected by extending the time $\eta$ into the complex plane by judiciously introducing a small imaginary component. In other words, we define the closed time integration contour in the complex $\eta$ plane, as shown in Fig.~\ref{tcontour}. The evolution starts at $\eta=-\infty(1-i\epsilon)$, progresses forward to the observation time $\eta_0$ (via contour $\mathcal{C}_-$), and then returns back to $\eta=-\infty(1+i\epsilon)$ (via contour $\mathcal{C}_+$). The tiny imaginary shifts $\pm\, i\epsilon$ damp the high‑frequency oscillations of early‑time modes and project the system onto the Bunch-Davies vacuum. It has been observed that it is always preferable to maintain this explicit initial-state perception in the computation to avoid the numerous complications that arise in the calculation at two vertices and beyond two-vertex calculations \cite{Adshead:2009cb}. It turns out to be extremely useful to employ the elegant diagrammatic approach to systematically arrange the extensive calculation. In this work, we shall employ the cosmological diagrammatic rules that were particularly developed to compute the in-in correlators \cite{Giddings:2010ui}. These diagrammatic rules are briefly summarised in the Appendix \ref{CDrules}. 
\subsection{Cubic order interaction Hamiltonian with inflationary metric perturbations}

Our aim is to calculate the 4-point autocorrelation functions of gauge fields that arise from the exchange interaction of metric fluctuations, i.e., scalar and tensor perturbations.
In other words, we are interested in the operator $\mathcal{O}$ as $A_iA_jA_mA_n$ or $A'_iA'_jA'_mA'_n$ in eq. (\ref{exp1_in-in}), which first requires the interaction Hamiltonian $H_{\text{int}}$ at the cubic order. Note that the gauge fields lack self-interactions thus the inflaton and gravity mediate all the interactions. Therefore, there are no contact diagrams in the perturbative calculation of such $n$-point gauge field autocorrelations that usually result from self-interactions. However, these gauge fields can interact with themselves through the exchange of the inflaton fluctuation or a graviton. By choosing to work in the comoving gauge where $\delta\phi=0$, one can remove these explicit inflaton fluctuations. However, such degrees of freedom manifest as a dynamical scalar degree of freedom in the metric fluctuations. Consequently, in the comoving gauge, the interactions of the gauge field are purely mediated by the gravitational sector which can be obtained by expanding the action in eq. (\ref{eq: EM action}) in the metric perturbations and the gauge field $A_i$. This yields the leading-order contribution at the cubic order, arising from the combination of one metric perturbation and a pair of gauge fields.
To achieve this, we treat the gravitational field as a Hamiltonian system, which is suitable for canonical quantization and utilize the Arnowitt–Deser–Misner (ADM) decomposition of the spacetime metric that splits the spacetime into a series of three-dimensional spatial slices that evolve over time. In this case, the spacetime metric is given by,
\begin{eqnarray} 
 ds^2=-N^2dt^2+h_{ij}(dx^i+N^i dt)(dx^j+N^jdt)~, 
 \end{eqnarray} 
  where $N(t,{\bf x})$ and $N^i(t,{\bf x})$ are called the lapse function and the shift vector, respectively. They are not dynamical variables but instead serve as Lagrange multipliers which are determined through the constraint equations. The dynamical degrees of freedom are contained in the spatial part of the metric $h_{ij }$ and is parameterized as  $h_{ij}=a^2 e^{2\zeta}[e^\gamma]_{ij}$ where $\zeta$ and $\gamma_{ij}$ are the curvature perturbation and tensor perturbation defined in section \ref{eq:mps}. By solving the constraint equations, we can express the lapse function and shift vector in terms of the spatial metric tensor $h_{ij }$. Subsequently, solving the constraint equations at the first order provides us with
\beq
N &=& 1 + \frac{\dot{\zeta}}{H}, \quad 
N_i = \partial_i\left(-\frac{\zeta}{H} + \epsilon\, a^2\, \partial^{-2} \dot{\zeta} \right)~, 
\eeq
 where an overdot denotes the time derivative with respect to cosmic time $t$ and $\partial^{-2}$ denotes the inverse Laplacian operator. Throughout this work, we will be working in the de Sitter limit of slow-roll inflation, i.e., $\epsilon \to 0$. In this limit, we obtain
 \beq
 \label{SRLapShif}
N = 1 + \partial_t\left(\frac{{\zeta}}{H}\right)+\mathcal{O}(\epsilon), \quad 
N_i = -\partial_i\left(\frac{\zeta}{H} \right)+ \mathcal{O}(\epsilon)~. 
\eeq
Since we neglect the interaction terms that are slow-roll suppressed, we can proceed with the above forms of the lapse function and the shift vector.

Let us now commence from the action in eq.~(\ref{eq: EM action}) and compute the interaction part by substituting the scalar metric perturbation up to the leading order using eq.~(\ref{SRLapShif}) and the spatial component of the metric. The resulting total Lagrangian can be divided into a free part and an interaction part. As mentioned earlier, the leading-order interaction Lagrangian $L_{\text{int}}$ would be at the cubic order in metric perturbations and gauge fields. It will have two terms: the first involves a scalar perturbation coupled to a pair of gauge fields, while the second involves a tensor perturbation coupled to two pairs of gauge fields. Up to the cubic order, one can use, $H_{\text{int}}=-L_{\text{int}}$. This allows us to express the cubic order interaction Hamiltonian, denoted as $H_{\text{int}}^{(3)}$, as a sum of the Hamiltonians that are responsible for the scalar and tensor exchange interactions, i.e., $H_{\text{int}}^{(3)} = H_{\zeta AA} + H_{\gamma AA}$.  This derivation of the interaction Hamiltonian at the cubic order has been outlined elsewhere \cite{Caldwell:2011ra, Jain:2012ga, Jain:2012vm, Motta:2012rn, Shiraishi:2012xt}. Thus, the cubic interaction Hamiltonian with curvature perturbation can be obtained as
\beq\label{Hint}
H_{\zeta AA} = \int d^3 x \left(a\dot\lambda\frac{1}{H}\zeta\left(\frac{1}{2}\dot A_i^2 -\frac{1}{4 a^2}F_{ij}^{2}\right)+\p_t\left(\mathcal{B}_{\zeta AA}\right)\right)+\mathcal{O}(\epsilon)~,
\eeq
where $i$, $j$ indicate spatial indices, $\dot{A}_{i}^{2}$ denotes $ \sum_i\dot{A}_i \dot{A}_i$ and $F^2_{ij}$ represents $\sum_{i,j}F_{ij}F_{ij}$. To derive the above Hamiltonian, we have eliminated the cross terms involving the time and space derivatives of the gauge field by employing the equation of motion for the gauge field. This would leave us with only the above form with both temporal and spatial boundary terms. In the above equation, we have discarded the spatial boundary terms because the spatial and temporal derivatives of the gauge fields vanish at spatial infinity. The remaining temporal boundary term is 
 \beq
 \mathcal{B}_{\zeta AA}=\frac{a \lambda}{2 H}\zeta \left(\dot A_i^2 +\frac{1}{2 a^2}F_{ij}^{2}\right).
\eeq
 At this stage, we focus on the bulk interaction and neglect the temporal boundary terms. In the standard framework, such boundary terms do contribute to the four-point exchange diagrams. These contributions can be classified into two categories: those originating solely from the boundary Hamiltonian and those arising from mixed terms involving the boundary and bulk Hamiltonians, as given in eq. (2.23) of \cite{Braglia:2024zsl}. We have explicitly evaluated these contributions using the cosmological diagrammatic techniques, without resorting to field redefinitions. Our analysis shows that all these contributions from the boundary Hamiltonian are proportional to the time derivative of the gauge field power spectrum. In our setup, for $n>1/2$, the gauge field mode function freezes in the super-Hubble regime. As a result, the boundary induced contributions are suppressed relative to those from the  bulk Hamiltonian. Physically, this suppression can be attributed to the redshifting of the boundary effect on the final time slice -- the end of inflation surface -- where its contribution to the trispectrum effectively becomes negligible compared to the bulk exchange contribution. Therefore, in what follows, we restrict our attention to the contributions arising from the bulk Hamiltonian.

From the interaction Hamiltonian in eq.~(\ref{Hint}), it is evident that the vertices are either proportional to the derivative of $\lambda$ or to the first slow-roll parameter. 
One can make an observation that the dynamical nature of the Maxwell-scalar coupling and the Hubble parameter are responsible for producing such an interaction. This is the same expression given in
\cite{Jain:2012ga}. Converting $H_{\zeta AA}$ into the conformal time coordinate and dropping the temporal boundary term gives, 
\beq
\label{WorkingHZAA}
H_{\zeta AA}\= n\int d^{3}x\; \lambda\zeta \l A_{i}'^{2} -\frac{1}{2}F_{ij}^{2}\r~.
\eeq
In addition, we have also used the parametric form of the dilatonic coupling as $\lambda\propto a^{2n}$. 
This interaction leads to the exchange of scalar perturbation.
  The above expression is the same as the interaction Hamiltonian given in eq. (46) of \cite{Motta:2012rn}. However, it differs in form from the one in eq. (3.15) of \cite{Jain:2012ga}. Despite this difference, both expressions are equivalent to boundary terms.
Similarly, we get the cubic Hamiltonian involving tensor perturbation as follows
\begin{equation}
\label{H_gAA}
             H_{\gamma AA}=\frac{1}{2} \int d^3x \, \lambda \,\gamma_{ij} \left(A'_iA'_j-(\partial_iA_l\partial_jA_l+\partial_lA_i\partial_lA_j)+2\partial_iA_l\partial_lA_j \right).
         \end{equation}
In this expression, we also follow the same conversion for the repeated indices summed over, irrespective of whether they are upper or lower. It is evident that the forms of $H_{\zeta AA}$ and $H_{\gamma AA}$ are similar. In both, the interaction vertex contains the same time-dependent coupling function $\lambda(\eta)$. They are proportional to the metric perturbation and the quadratic form of the electromagnetic field. The electromagnetic part is enclosed in a bracket, which contains the first term from the electric part of the action and the second term from the magnetic part of the action. 

\section{Inflationary trispectrum of gauge fields with exchange interactions}
\label{section:4}
As discussed earlier, gauge fields interact via the exchange of metric fluctuations.
In this section, we implement the interaction Hamiltonian derived in the preceding section in the in-in formalism to compute the relevant higher-order correlators.
In particular, our aim is to calculate the four-point autocorrelation function for the gauge field, expressed in terms of its associated EM fields as $\langle B_\mu B_\nu B_\rho B_\sigma\rangle$ and $\langle E_\mu E_\nu E_\rho E_\sigma\rangle$, where $E_\mu$ and $B_\mu$ are the associated electric and magnetic fields defined in section \ref{eq:mps}, and they are manifestly invariant under EM gauge transformations. Recall that, we have been working in the Coulomb gauge with $A_0=0$ and $\partial_iA_i=0$. Since $A_\mu$ is not gauge invariant, we first compute the four-point functions $\langle A_iA_jA_mA_n\rangle$ and $\langle A’_iA’_jA’_mA’_n\rangle$ in the Coulomb gauge which enables us to obtain the desired autocorrelations in terms of the electric and magnetic fields. Moreover, we explore the features of such correlations for different choices of the power law index $n$ of coupling function, say $n=0,1,2,3$, and discuss  some specific cases, highlighting their growing behaviors. 

\subsection{Scalar exchanges}

We begin by evaluating the scalar-exchange contribution to the correlator. Making use of the interaction Hamiltonian
$H_{\zeta AA}$ in eq. (\ref{WorkingHZAA}), we apply the in-in master formula to compute the leading-order four-point function
$\expval{A_{i}\l \vb{k}_{1},\eta_0\r A_{j}\l \vb{k}_{2},\eta_0\r A_{m}\l \vb{k}_{3},\eta_0\r A_{n}\l \vb{k}_{4},\eta_0\r}$ in Coulomb gauge. 
The non-trivial contribution arises at quadratic order in the perturbative expansion of eq. (\ref{exp1_in-in}), corresponding to the exchange of a curvature perturbation $\zeta$. Since the perturbative expansion generates a large number of terms, a systematic organisation is required. In this regard, the cosmological diagrammatic approach provides a convenient and intuitive framework, facilitating efficient accounting for both time-ordered and anti-time-ordered contributions arising from the interaction Hamiltonian.
The corresponding interaction vertex is represented by a three-leg diagram, as shown in Fig.~\ref{vertex}. The two solid legs denote gauge fields, while the remaining leg represents a metric fluctuation (either a curvature perturbation or a graviton).
With four external legs and the interaction vertex, we can construct all the relevant diagrams. 
The structure of the vertex indicates that the leading-order four-point gauge field correlator arises from exchange diagrams.
Thus, the connected four-point function constitutes a genuine non-Gaussian signal sourced by the exchange of metric fluctuations.

\begin{figure}
\centering
\begin{tikzpicture}
\begin{scope}[scale=2.5]
  \coordinate (v1) at (-0.32, 0);\fill (v1) circle (0.7pt);
  \coordinate (v2) at (0.35,0);\draw[fill] (v2);
  \coordinate (u1) at (-0.8,0.5);
  \coordinate (u2) at (-0.8,-0.5);
  \node[above=1pt] at (u1) {$\text{A}_{i}$};
  \node[below=1pt] at (u2) {$\text{A}_{j}$};
  \draw[black,line width=1.2pt] (v1) -- (u1);
  \draw[black,line width=1.2pt] (v1) -- ( u2);
  \draw[decorate, decoration={snake, amplitude=1mm, segment length=3mm},line width=1.2pt,gray] (v1) -- ( v2);
b  \node[below=2pt] at (0,0) {$\mathbb{\zeta}$/$\mathbb{\gamma}$};
\coordinate (v1) at (-0.32, 0);\fill (v1) circle (0.7pt);
\end{scope}
\end{tikzpicture} 
\caption{The interaction vertex corresponding to the Hamiltonian in eqs. (\ref{WorkingHZAA}) and (\ref{H_gAA}). The solid lines denote the gauge field legs while the wavy line represents the metric fluctuations $\zeta$ or $\gamma$, respectively.}
\label{vertex}
\end{figure}
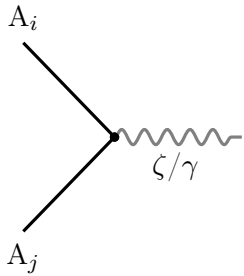

\begin{figure}[t]
\centering
\begin{tikzpicture}
\begin{scope}[scale=2.5]
  \draw[dashed] (-0.9,0.8) -- (0.9,0.8);
  \coordinate (u1) at (-0.8,0.8);
  \coordinate (u2) at (-0.3,0.8);
  \coordinate (u3) at (0.3,0.8);
  \coordinate (u4) at (0.8,0.8);
  \node[above=1pt] at (u1) {\small $\text{A}_{i}(\vb{k}_{1})$};
  \node[above=1pt] at (u2) {\small $\text{A}_{j}(\vb{k}_{2})$};
  \node[above=1pt] at (u3) {\small $\text{A}_{l}(\vb{k}_{3})$};
  \node[above=1pt] at (u4) {\small $\text{A}_{m}(\vb{k}_{4})$};
  \coordinate (v1) at (-0.5,0.2);\fill (v1) circle (0.7pt);
  \node[below=1pt] at (v1) {\small $-i2n\lambda$};
  \coordinate (v2) at (0.5,0.2);\draw[fill] (v2) circle (0.7pt);
  \node[below=1pt] at (v2) {\small $-i2n\lambda$};
  \draw[black,line width=1pt] (v1) -- (-0.8,0.8);
  \draw[black,line width=1pt] (v1) -- ( -0.3,0.8);
  \draw[black,line width=1pt] (v2) -- ( 0.3,0.8);
  \draw[black,line width=1pt] (v2) -- ( 0.8,0.8);
  \draw[decorate, decoration={snake, amplitude=0.6mm, segment length=2mm},gray,Black] (v1) -- ( v2);
\end{scope}
\end{tikzpicture}
\begin{tikzpicture}
\begin{scope}[scale=2.5]
  \draw[dashed] (-0.9,0.8) -- (0.9,0.8);
  \coordinate (v1) at (-0.32, 1.4);\fill (v1) circle (0.7pt);
  \node[above=1pt] at (v1) {\small $i2n\lambda$};
  \coordinate (v2) at (0.35,0.2);\draw[fill] (v2) circle (0.7pt);
  \node[below=1pt] at (v2) {\small $-i2n\lambda$};
  \coordinate (u1) at (-0.8,0.8);
  \coordinate (u2) at (-0.3,0.8);
  \coordinate (u3) at (0.3,0.8);
  \coordinate (u4) at (0.8,0.8);
  \node[below=1pt] at (u1) {\small $\text{A}_{i}(\vb{k}_{1})$};
  \node[below=1pt] at (u2) {\small $\text{A}_{j}(\vb{k}_{2})$};
  \node[above=1pt] at (u3) {\small $\text{A}_{l}(\vb{k}_{3})$};
  \node[above=1pt] at (u4) {\small $\text{A}_{m}(\vb{k}_{4})$};
  \draw[black,line width=1pt] (v1) -- (-.8,0.8);
  \draw[black,line width=1pt] (v1) -- ( -0.35,0.8);
  \draw[black,line width=1pt] (v2) -- ( 0.35,0.8);
  \draw[black,line width=1pt] (v2) -- ( 0.8,0.8);
  \draw[decorate, decoration={snake, amplitude=0.6mm, segment length=2mm},thick,gray] (v1) -- ( v2);
\end{scope}
\end{tikzpicture} 
\begin{tikzpicture}
\begin{scope}[scale=2.5]
  \draw[dashed] (-0.9,0.8) -- (0.9,0.8);
  \coordinate (u1) at (-0.8,0.8);
  \coordinate (u2) at (-0.3,0.8);
  \coordinate (u3) at (0.3,0.8);
  \coordinate (u4) at (0.8,0.8);
  \node[above=1pt] at (u1) {\small $\text{A}_{i}(\vb{k}_{1})$};
  \node[above=1pt] at (u2) {\small $\text{A}_{j}(\vb{k}_{2})$};
  \node[above=1pt] at (u3) {\small $\text{A}_{l}(\vb{k}_{3})$};
  \node[above=1pt] at (u4) {\small $\text{A}_{m}(\vb{k}_{4})$};
  \coordinate (v1) at (-0.5,0.2);\fill (v1) circle (0.7pt);
  \node[below=1pt] at (v1) {\small $-i2n\lambda$};
  \coordinate (v2) at (0.5,0.2);\draw[fill] (v2) circle (0.7pt);
  \node[below=1pt] at (v2) {\small $-i2n\lambda$};
  \draw[black,line width=1pt] (v1) -- (-0.8,0.8);
  \draw[black,line width=1pt] (v2) -- ( 0.05,0.51);
  \draw[black,line width=1pt] (0.05,0.5) arc[start angle=-30,end angle=130,radius=0.08cm];
  \draw[black,line width=1pt] (-0.06,0.6) -- (-0.3,0.8);
  \draw[black,line width=1pt] (v1) -- ( 0.3,0.8);
  \draw[black,line width=1pt] (v2) -- ( 0.8,0.8);
  \draw[decorate, decoration={snake, amplitude=0.6mm, segment length=2mm},thick,gray] (v1) -- ( v2);
  \vspace{1cm}
\end{scope}
\end{tikzpicture}
\begin{tikzpicture}
\begin{scope}[scale=2.5]
  \draw[dashed] (-0.9,0.8) -- (0.9,0.8);
  \coordinate (v1) at (-0.1, 1.4);\fill (v1) circle (0.7pt);
  \node[above=1pt] at (v1) {\small $i2n\lambda$};
  \coordinate (v2) at (0.1,0.2);\draw[fill] (v2) circle (0.7pt);
  \node[below=1pt] at (v2) {\small $-i2n\lambda$};
  \coordinate (u1) at (-0.8,0.8);
  \coordinate (u2) at (-0.3,0.8);
  \coordinate (u3) at (0.3,0.8);
  \coordinate (u4) at (0.8,0.8);
  \node[below=1pt] at (u1) {\small $\text{A}_{i}(\vb{k}_{1})$};
  \node[above=1pt] at (u2) {\small $\text{A}_{j}(\vb{k}_{2})$};
  \node[below=1pt] at (u3) {\small $\text{A}_{l}(\vb{k}_{3})$};
  \node[above=1pt] at (u4) {\small $\text{A}_{m}(\vb{k}_{4})$};
  \draw[black,line width=1pt] (v1) -- (-.8,0.8);
  \draw[black,line width=1pt] (v2) -- ( -0.35,0.8);
  \draw[black,line width=1pt] (v1) -- ( 0.35,0.8);
  \draw[black,line width=1pt] (v2) -- ( 0.8,0.8);
  \draw[decorate, decoration={snake, amplitude=0.6mm, segment length=2mm},thick,gray] (v1) -- ( v2);
\end{scope}
\end{tikzpicture} 
\begin{tikzpicture}
\begin{scope}[scale=2.5]
  \draw[dashed] (-0.9,0.8) -- (0.9,0.8);
  \coordinate (u1) at (-0.8,0.8);
  \coordinate (u2) at (-0.3,0.8);
  \coordinate (u3) at (0.3,0.8);
  \coordinate (u4) at (0.8,0.8);
  \node[above=1pt] at (u1) {\small $\text{A}_{i}(\vb{k}_{1})$};
  \node[above=1pt] at (u2) {\small $\text{A}_{j}(\vb{k}_{2})$};
  \node[above=1pt] at (u3) {\small $\text{A}_{l}(\vb{k}_{3})$};
  \node[above=1pt] at (u4) {\small $\text{A}_{m}(\vb{k}_{4})$};
  \coordinate (v1) at (0,0.2);\fill (v1) circle (0.7pt);
  \node[below=1pt] at (v1) {\small $-i2n\lambda$};
  \coordinate (v2) at (0,0.6);\draw[fill] (v2) circle (0.7pt);
  \node[right=1pt] at (v2) {\small $-i2n\lambda$};
  \draw[black,line width=1pt] (v1) -- (-0.8,0.8);
  \draw[black,line width=1pt] (v2) -- ( -0.3,0.8);
  \draw[black,line width=1pt] (v2) -- ( 0.3,0.8);
  \draw[black,line width=1pt] (v1) -- ( 0.8,0.8);
  \draw[decorate, decoration={snake, amplitude=0.6mm, segment length=2mm},thick,gray] (v1) -- ( v2);
\end{scope}
\end{tikzpicture}
\begin{tikzpicture}
\begin{scope}[scale=2.5]
  \draw[dashed] (-0.9,0.8) -- (0.9,0.8);
  \coordinate (v1) at (0, 1.4);\fill (v1) circle (0.7pt);
  \node[above=1pt] at (v1) {\small$i2n\lambda$};
  \coordinate (v2) at (0,0.2);\draw[fill] (v2) circle (0.7pt);
  \node[below=1pt] at (v2) {\small $-i2n\lambda$};
  \coordinate (u1) at (-0.8,0.8);
  \coordinate (u2) at (-0.3,0.8);
  \coordinate (u3) at (0.3,0.8);
  \coordinate (u4) at (0.8,0.8);
  \node[below=1pt] at (u1) {\small$\text{A}_{i}(\vb{k}_{1})$};
  \node[above=1pt] at (u2) {\small$\text{A}_{j}(\vb{k}_{2})$};
  \node[above=1pt] at (u3) {\small\small$\text{A}_{l}(\vb{k}_{3})$};
  \node[below=1pt] at (u4) {\small$\text{A}_{m}(\vb{k}_{4})$};
  \draw[black,line width=1pt] (v1) -- (-.8,0.8);
  \draw[black,line width=1pt] (v2) -- ( -0.35,0.8);
  \draw[black,line width=1pt] (v2) -- ( 0.35,0.8);
  \draw[black,line width=1pt] (v1) -- ( 0.8,0.8);
  \draw[decorate, decoration={snake, amplitude=0.6mm, segment length=2mm},thick, gray] (v1) -- ( v2);
\end{scope}
\end{tikzpicture} 
\caption{Scalar-exchange diagrams contributing to the connected four-point correlator 
$\langle A_{i}A_{j}A_{l}A_{m} \rangle$ of gauge fields $A_i(\vb{k})$. 
Each cubic vertex represents the interaction eq. (\ref{WorkingHZAA}), involving two gauge field modes 
$A_i(\vb{k})$ and the exchanged scalar $\zeta$. 
External legs are labeled by their comoving momenta $\boldsymbol{k}_i$, with overall momentum conservation 
$\sum^4_{i=1} \boldsymbol{k}_i = 0$.  The dashed horizontal line indicates the final time hypersurface ($\eta=\eta_0$), 
on which all external operators are evaluated. The rows are arranged in a manner that allows three distinct ways to connect external legs via an intermediate propagator. We call each row a distinctive channel.
We label each channel with the corresponding contraction. The top diagram is called the $(12)(34)$ channel contribution. Middle one is  $(13)(24)$  and bottom one is $(14)(23)$.}
\label{diagram}
\end{figure}
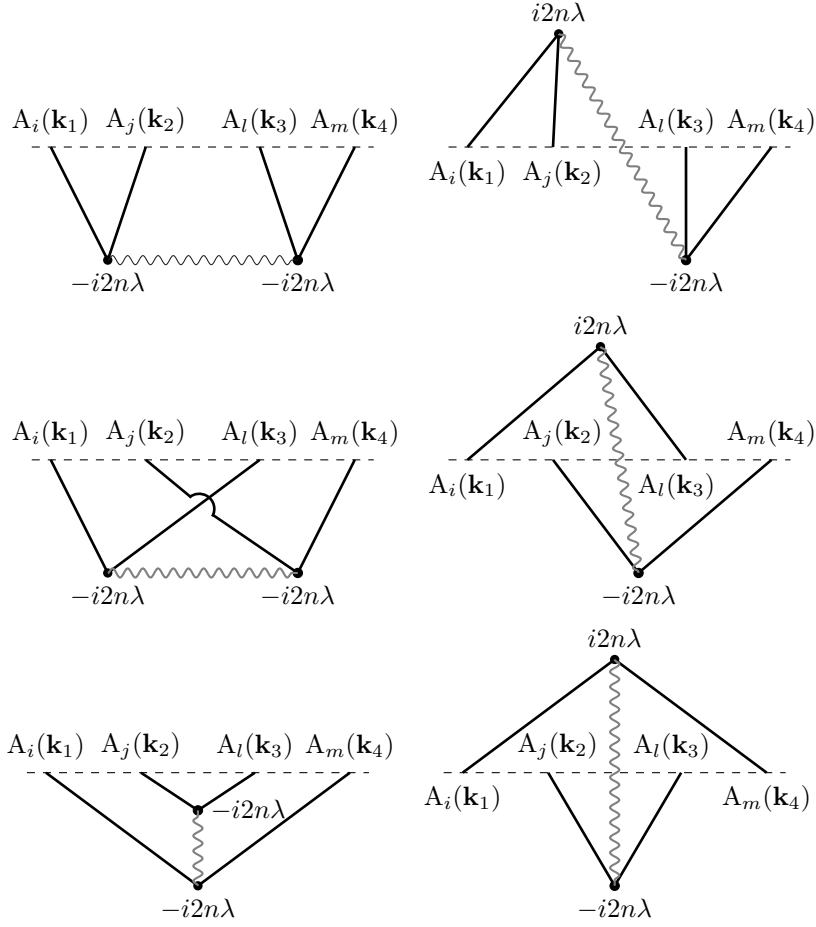
Applying the cosmological diagrammatic rules outlined in Appendix~\ref{CDrules}, we construct all possible diagrams for scalar-exchange interactions.
In Fig.~\ref{diagram}, we present all six possible diagrams, which should also be accompanied by their mirror reflections about the final time surface $\eta_0$, indicated by the dashed line in the figure. Due to the double exponential structure in the in-in expression, vertices can originate from either the forward contour or the backward contour. When translating these diagrams into their corresponding mathematical expressions using the diagrammatic rules, we meticulously record the time contour by assigning a label $(+)$ to the vertices of the forward branch $\mathcal{C_+}$ and a $(-) $ to those of the backward branch $\mathcal{C_-}$. In all diagrams, the vertices below the evaluation time surface $\eta_0$ are in the $\mathcal{C_+}$ contour while the vertices above the $\eta_0$ are in $\mathcal{C_-}$. For instance, diagrams on the left side are from vertices $(++)$, while their mirror reflections are from $(--)$ vertices. Similarly, diagrams on the right side are labelled as $(-+)$, and their reflection contributions are labelled as $(+-)$. This method of tracking the contour will help avoid the occurrence of spurious divergences that may arise during the performance of such calculations by setting external momenta in specific choices \cite{Adshead:2009cb}. This enables us to cross check our results with the desired choices of external momenta. 
To evaluate the correlator, we must sum all these twelve diagrams, where the mirror reflections correspond to complex conjugates. Consequently, we only need to add these six diagrams and extract the two-time real part of their sum. It is further evident that we only need to consider two diagrams positioned at the top in Fig.~\ref{diagram}, and the contributions of the remaining diagrams can be readily obtained by permuting the external legs from the sum of the first two diagrams, i.e., 
\beq
\expval{A_{i}\l \vb{k}_{1}\r A_{j}\l \vb{k}_{2}\r A_{l}\l \vb{k}_{3}\r A_{m}\l \vb{k}_{4}\r} = 2\,\text{Re}\left\{
\tikz[baseline={(0,0.1)}]{
  \draw[dashed] (-0.9,0.8) -- (0.9,0.8);
  \coordinate (v1) at (-0.5,0.2);\fill (v1) circle (1pt);
  \coordinate (v2) at (0.5,0.2);\draw[fill] (v2) circle (1pt);
  \draw[black,line width=0.6pt] (v1) -- (-0.8,0.8);
  \draw[black,line width=0.6pt] (v1) -- ( -0.3,0.8);
  \draw[black,line width=0.6pt] (v2) -- ( 0.3,0.8);
  \draw[black,line width=0.6pt] (v2) -- ( 0.8,0.8);
  \draw[decorate, decoration={snake, amplitude=0.6mm, segment length=2mm},line width=0.6pt,gray] (v1) -- ( v2);
}\; 
+ \tikz[baseline={(0,0.7)}]{
  \draw[dashed] (-0.9,0.8) -- (0.9,0.8);
  \coordinate (v1) at (-0.32, 1.4);\fill (v1) circle (1pt);
  \coordinate (v2) at (0.35,0.2);\draw[fill] (v2) circle (1pt);
  \draw[black,line width=0.6pt] (v1) -- (-.8,0.8);
  \draw[black,line width=0.6pt] (v1) -- ( -0.35,0.8);
  \draw[black,line width=0.6pt] (v2) -- ( 0.35,0.8);
  \draw[black,line width=0.6pt] (v2) -- ( 0.8,0.8);
  \draw[decorate, decoration={snake, amplitude=0.6mm, segment length=2mm},line width =0.6pt,gray] (v1) -- ( v2);
}\; 
\right\}
+\text{permutations}~.
\eeq
From the Hamiltonian in eq. (\ref{WorkingHZAA}), it is clear that the final result for the four-point correlator would be proportional to $n^2$. Thus, with $\lambda \sim a^{2n}$, expressing $\dot{\lambda}/H\lambda =2n$ and using the diagrammatic rules, we obtain the final result as 
\beq\label{autocorr}
\expval{A_{i}\l \boldsymbol{k}_{1}\r A_{j}\l \boldsymbol{k}_{2}\r A_{l}\l \boldsymbol{k}_{3}\r A_{m}\l \boldsymbol{k}_{4}\r}\=  \l 2\pi\r^{3}\delta^{(3)}\l \sum_{\alpha=1}^4\boldsymbol{k}_\alpha\r \l\frac{\dot{\lambda}}{H \lambda}\r^2 \sum_{X,Y \in \{A,A’\}} Q^{\zeta XY}_{ijlm}\, I^{XY}\\
& + (j,\boldsymbol{k}_{2})\leftrightarrow (l,\boldsymbol{k}_{3})+ (j,\boldsymbol{k}_{2})\leftrightarrow (m,\boldsymbol{k}_{4})~,
\eeq
where $Q$'s are the polarisation factors which we express them explicitly below, and $I$ are from the in-in time integrals. The superscript $\zeta$ indicates that the calculation is performed for the scalar exchange, while the superscript $XY$ leads to four distinct combinations, as $X$ and $Y$ can take values from the sets $\{A, A’\}$. The rationale behind this notation comes from the structure of the interaction Hamiltonian, which contains two types of terms: those involving the time derivative of the gauge field and those involving its spatial derivative. In Fourier space, the corresponding contractions naturally give rise to Wightman functions built either from the gauge field mode function itself or from its time derivative. Before we move on specifying the integrals, let us express them in terms of different contour contributions as follows  
\begin{eqnarray}
\label{IAAdef}
I^{A'A'}&=& 2\text{Re}\ll I^{A'A'}_{++}+I^{A'A'}_{-+}\rr,\quad I^{AA}= 2\text{Re}\ll I^{AA}_{++}+I^{AA}_{-+}\rr,\\
I^{AA'}&=& 2\text{Re}\ll I^{AA'}_{++}+I^{AA'}_{-+}\rr,\quad I^{A'A}= 2\text{Re}\ll I^{A'A}_{++}+I^{A'A}_{-+}\rr~.
\label{IAA'def}
\end{eqnarray}
For brevity, we can use the above introduced notation to express these expressions as $I^{XY} = 2\text{Re}\ll I^{XY}_{++} + I^{XY}_{-+}\rr$, with $X,Y \in \{A,A’\}$. Then the corresponding integrals are defined as 
\begin{eqnarray}
\label{Ipp1}
I^{XY}_{++}&=& -\int_{-\infty}^{\eta_{0}}d\eta_{1}\;\lambda\l \eta_{1}\r\int_{-\infty}^{\eta_{0}}d\eta_{2}\;\lambda\l \eta_{2}\r \nonumber \\
&& W_{k_{1}}^{AX}\l \eta_{0},\eta_{1}\r W_{k_{2}}^{AX}\l \eta_{0},\eta_{1}\r G_{k_{\mathrm{I}}}^{\zeta}\l \eta_{1},\eta_{2}\r W_{k_{3}}^{AY}\l \eta_{0},\eta_{2}\r W_{k_{4}}^{AY}\l \eta_{0},\eta_{2}\r,\\
I^{XY}_{-+} &=& \int_{-\infty}^{\eta_{0}}d\eta_{1}\;\lambda\l \eta_{1}\r\int_{-\infty}^{\eta_{0}}d\eta_{2}\;\lambda\l \eta_{2}\r \nonumber \\
&& W_{k_{1}}^{XA}\l \eta_{1},\eta_{0}\r W_{k_{2}}^{XA}\l \eta_{1},\eta_{0}\r\wc_{k_{\mathrm{I}}}\l \eta_{1},\eta_{2}\r W_{k_{3}}^{AY}\l \eta_{0},\eta_{2}\r W_{k_{4}}^{AY}\l \eta_{0},\eta_{2}\r,
\label{Imp1}
\end{eqnarray}
where, we have introduced the following Wightman functions:
$W^{XY}_k(\eta_1, \eta_2) = X_k(\eta_1) Y^{*}_k(\eta_2)$ and 
$W^{\zeta}_k(\eta_1, \eta_2) = \zeta_k(\eta_1) \zeta^{*}_k(\eta_2)$.  In addition, the above expression also involves the Feynman propagator, which is defined as 
\begin{eqnarray}
G_k^\zeta( \eta_1, \eta_2) = W_k^\zeta(\eta_1, \eta_2) \, \theta(\eta_1 - \eta_2) + W_k^\zeta(\eta_2, \eta_1) \, \theta(\eta_2-\eta_1)\;.
\end{eqnarray}

Note that, the propagator $G^{\zeta}_{k_{\mathrm{I}}}$ and Wightman functions $W^{\zeta}_{k_{\mathrm{I}}}$ characterize the exchange of the curvature perturbation, where $k_{\mathrm{I}}$ is the momentum of the propagator, given by the magnitude of the vector $\boldsymbol{k}_{\mathrm{I}}=\boldsymbol{k}_1+\boldsymbol{k}_2=-\l\boldsymbol{k}_3+\boldsymbol{k}_4\r$ from the momentum conservation at each vertex. With these definitions, the integrals $I$’s appearing in eq. (\ref{autocorr}) can be evaluated once the mode functions of gauge fields and curvature perturbations are specified. 
We now explicitly express the polarisation coefficient in front of each integral in eq. (\ref{autocorr}) which can be compactly written in terms of the projection matrix $\Pi_{ij}$ and generalized projection matrix $\mathbb{P}_{ij}$, given by 
\begin{eqnarray}            
Q^{\zeta AA}_{ijlm} &=& \Pi_{ia}\!\left( \hat{\boldsymbol{k}}_{1}\right) \mathbb{P}_{ab}\!\left( \boldsymbol{k}_{2}, \boldsymbol{k}_{1}\right)
               \Pi_{bj}\!\left( \hat{\boldsymbol{k}}_{2}\right) \; 
               \Pi_{lc}\!\left( \hat{\boldsymbol{k}}_{3}\right) \mathbb{P}_{cd}\!\left( \boldsymbol{k}_{4}, \boldsymbol{k}_{3}\right)
               \Pi_{dm}\!\left( \hat{\boldsymbol{k}}_{4}\right),\label{QA2} \\               
Q^{\zeta AA'}_{ijlm} &=& \Pi_{ia}\!\left( \hat{\boldsymbol{k}}_{1}\right)\mathbb{P}_{ab}\!\left( \boldsymbol{k}_{2}, \boldsymbol{k}_{1}\right) 
               \Pi_{bj}\!\left( \hat{\boldsymbol{k}}_{2}\right) \;
               \Pi_{lc}\!\left( \hat{\boldsymbol{k}}_{3}\right) 
               \Pi_{cm}\!\left( \hat{\boldsymbol{k}}_{4}\right), \label{QA3}\\        Q^{\zeta A'A}_{ijlm} &=& \Pi_{ia}\!\left( \hat{\boldsymbol{k}}_{1}\right) 
               \Pi_{aj}\!\left( \hat{\boldsymbol{k}}_{2}\right) \;
               \Pi_{lc}\!\left( \hat{\boldsymbol{k}}_{3}\right)\mathbb{P}_{cd}\!\left( \boldsymbol{k}_{4}, \boldsymbol{k}_{3}\right)  
               \Pi_{dm}\!\left( \hat{\boldsymbol{k}}_{4}\right),\label{QA4} \\               
Q^{\zeta A'A'}_{ijlm} &=& \Pi_{ia}\!\left( \hat{\boldsymbol{k}}_{1}\right) 
               \Pi_{aj}\!\left( \hat{\boldsymbol{k}}_{2}\right) \;
               \Pi_{lc}\!\left( \hat{\boldsymbol{k}}_{3}\right) 
               \Pi_{cm}\!\left( \hat{\boldsymbol{k}}_{4}\right),\label{QA1}
\end{eqnarray}
where $\Pi_{ij}\!\left( \hat{\boldsymbol{k}}\right)$ is the standard projection matrix onto the plane that is orthogonal to the direction of $\boldsymbol{k}$, is defined as $\Pi_{ij}\!\left( \hat{\boldsymbol{k}}\right)= \delta_{i j}-\hat{k}_{i}\hat{k}_{j}$, and $\mathbb{P}_{ab}\!\left( \boldsymbol{k}_{1}, \boldsymbol{k}_{2}\right)$ is the bilinear generalization of the projection matrix defined as, 
\begin{eqnarray}
    \mathbb{P}_{ab}\!\left( \boldsymbol{k}_{1}, \boldsymbol{k}_{2}\right)= \l\boldsymbol{k}_{1}\cdot\boldsymbol{k}_{2}\r\delta_{ab}-k_{1a}k_{2b}~.
\end{eqnarray}
If the arguments of the generalized projection matrix are the same vector, it will reduce to the standard projection matrix, i.e., $\mathbb{P}_{ab}\!\left( \boldsymbol{k}, \boldsymbol{k}\right)=k^2\,\Pi_{ab}\!\left( \hat{\boldsymbol{k}}\right)$. Thus, expressing the polarisation coefficients of eq.~(\ref{autocorr}) as above makes it evident that the resulting correlator satisfies the Coulomb gauge condition. Contracting any external gauge field momentum with its free index yields zero, showing explicitly that the polarisation coefficient enforces the EM gauge constraints. For example, if we contract $k_{1i}$ with our correlator in eq.~(\ref{autocorr}), it will manifestly yield zero. This is because it is trivial to observe that $k_{1i}$ will be contracted with four polarisation coefficients, and for each polarisation coefficient, it will be contracted with the projection matrix $\Pi_{ia}\left(\hat{\boldsymbol{k}}_1\right)$ in all $Q$'s, which is identically zero. Similarly, it is also important to note that $\mathbb{P}_{ab}\!\left( \boldsymbol{k}_{1}, \boldsymbol{k}_{2}\right)k_{1b}=0$ and $\mathbb{P}_{ab}\!\left( \boldsymbol{k}_{1}, \boldsymbol{k}_{2}\right)k_{2a}=0$. From these properties of $\mathbb{P}$, one finds that 
\beq
    \Pi_{ia}\!\left( \hat{\boldsymbol{k}}_{1}\right) \mathbb{P}_{ab}\!\left( \boldsymbol{k}_{2}, \boldsymbol{k}_{1}\right) 
               \Pi_{bj}\!\left( \hat{\boldsymbol{k}}_{2}\right) \, &=\mathbb{P}_{ij}\!\left( \boldsymbol{k}_{2}, \boldsymbol{k}_{1}\right).
\eeq
Therefore, the expressions in eqs. (\ref{QA2})–(\ref{QA1}) take the form, 
\begin{eqnarray}
\label{fQA2}               
Q^{\zeta AA}_{ijlm} &=& \mathbb{P}_{ij}\!\left( \boldsymbol{k}_{2}, \boldsymbol{k}_{1}\right)
                \; 
               \mathbb{P}_{lm}\!\left( \boldsymbol{k}_{4}, \boldsymbol{k}_{3}\right),\\
\label{fQA3}               
Q^{\zeta AA'}_{ijlm} &=& \mathbb{P}_{ij}\!\left( \boldsymbol{k}_{2}, \boldsymbol{k}_{1}\right) \;
               \Pi_{lc}\!\left( \hat{\boldsymbol{k}}_{3}\right) 
               \Pi_{cm}\!\left( \hat{\boldsymbol{k}}_{4}\right), 
                \\
\label{fQA4}               
Q^{\zeta A'A}_{ijlm} &=& \Pi_{ia}\!\left( \hat{\boldsymbol{k}}_{1}\right) 
               \Pi_{aj}\!\left( \hat{\boldsymbol{k}}_{2}\right) \;
              \mathbb{P}_{lm}\!\left( \boldsymbol{k}_{4}, \boldsymbol{k}_{3}\right),\\
              \label{fQA1}
Q^{\zeta A'A'}_{ijlm} &=& \Pi_{ia}\!\left( \hat{\boldsymbol{k}}_{1}\right) 
               \Pi_{aj}\!\left( \hat{\boldsymbol{k}}_{2}\right) \;
               \Pi_{lc}\!\left( \hat{\boldsymbol{k}}_{3}\right) 
               \Pi_{cm}\!\left( \hat{\boldsymbol{k}}_{4}\right).
\end{eqnarray}
In principle, one can further eliminate the dummy indices in the above expressions by performing their contraction, thereby obtaining the polarisation coefficient solely in terms of free indices. To do so, we only need to compute the following two terms
\begin{eqnarray}
\Pi_{ia}\!\left( \hat{\boldsymbol{k}}_{1}\right) 
               \Pi_{aj}\!\left( \hat{\boldsymbol{k}}_{2}\right)&=&\delta_{ij} -\hat{k}_{1i}\hat{k}_{1j}-\hat{k}_{2i}\hat{k}_{2j}+ \l \hat{\boldsymbol{k}}_{1}\cdot\hat{\boldsymbol{k}}_{2}\r \hat{k}_{1i}\hat{k}_{2j}.
\end{eqnarray}
These results are sufficient to extrapolate the other free indices part, i.e., we just have to replace $i,j$ with $l,m$ and $\boldsymbol{k}_{1}, \boldsymbol{k}_{2}$ with $\boldsymbol{k}_{3}, \boldsymbol{k}_{4}$. However, for clarity, we retain the above form of the polarisation coefficients in eqs. (\ref{fQA2})–(\ref{fQA1}) so that its Coulomb gauge constraints are manifest. To this end, we have successfully shown that the four-point autocorrelation function $\langle A_iA_jA_lA_m\rangle$ takes the form as in eq. (\ref{autocorr}) and specified the necessary details to evaluate its right-hand side.
The next step is to evaluate these expressions for a specific choice of the coupling function. With the parametrisation $\lambda(a)\sim a^{2n}$, this amounts to fixing the value of $n$. This, in turn, determines the functional form of the gauge field mode function in eq. (\ref{modfn_A}), which can then be used to evaluate the integrals. Before proceeding with the specifics of the integral evaluation, let us define the correlator in a gauge-invariant manner. As previously mentioned, the gauge field correlator is not EM gauge-invariant and it would be more appropriate to express it in terms of the associated electric and magnetic fields. 

\begin{figure}[t]
\centering
\begin{tikzpicture}[scale=6, line join=round, >=latex]

  \coordinate (A) at (-0.7,0.4,-0.60);   
  \coordinate (B) at (0.15,.35, 0.7);   
  \coordinate (C) at ( 0.7,0.80, 0.55);   
  \coordinate (D) at (-0.3,0.15,-0.7);   

  \draw[black,->-=0.5,>=stealth,thick] (B) -- (A) node[pos=0.6,below=8pt] {$\vb{k}_1$};
 \draw[black,->-=0.5,>=stealth,thick] (C) -- (B) node[pos=0.3,below=8pt] {$\vb{k}_4$};
 \draw[black,->-=0.5,>=stealth,thick] (D)-- (C) node[pos=0.4,above=3pt] {$\vb{k}_3$};
 \draw[black,->-=0.5,>=stealth,thick] (A)--(D)node[pos=0.6,above=3pt] {$\vb{k}_2$};
 \draw[black,->-=0.5,>=stealth,thick] (B) -- (D)node[pos=0.6,left=1pt] {$\vb{k}_{I}$};
\end{tikzpicture}
\caption{A general representation of the momentum quadrilateral associated with the trispectrum. Note that, the two triangles need not lie in the same plane, and the total momentum conservation condition must always be satisfied.}
\label{quadrilateral}
\end{figure}
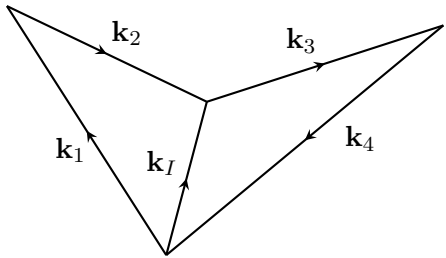

We first focus on the magnetic field autocorrelation function and translate our result to $\langle B_\mu B_\nu B_\rho B_\sigma\rangle$. However, for the clarity of presentation and phenomenological motivation, let us examine the correlator that does not have free indices, i.e., let's contract the first two magnetic fields and the last two, to write it as $\langle \bold{B}(\bold{k}_{1})\cdot\bold{B}(\bold{k}_{2})\; \bold{B}(\bold{k}_{3})\cdot \bold{B}(\bold{k}_{4})\rangle$. Since the magnetic field is defined with respect to a comoving observer, and the temporal component of the magnetic field vanishes in Coulomb gauge, i.e., $B_0=0$, we find that the gauge field $A_i$ can be related to the magnetic field as  
\begin{eqnarray}  B_\mu(\boldsymbol{k}_1)B^\mu(\boldsymbol{k}_2)=B_i(\boldsymbol{k}_1)B^i(\boldsymbol{k}_2)=-\frac{1}{a^4}\mathbb{P}_{ij}\!\left( \boldsymbol{k}_{2}, \boldsymbol{k}_{1}\right) A_i(\boldsymbol{k}_{1})A_j(\boldsymbol{k}_{2}).
\end{eqnarray}
Using this relation, the four–point function of the magnetic fields reduces to 
\begin{align}
\label{A2B}
 \langle B_\mu(\boldsymbol{k}_1)B^\mu(\boldsymbol{k}_2) B_\nu(\boldsymbol{k}_3)B^\nu(\boldsymbol{k}_4)\rangle &=\frac{1}{a^8}  \mathbb{P}_{ij}\!\left( \boldsymbol{k}_{2} , \boldsymbol{k}_{1}\right) \mathbb{P}_{lm}\!\left( \boldsymbol{k}_{4} , \boldsymbol{k}_{3}\right) \langle A_i(\boldsymbol{k}_{1})A_j(\boldsymbol{k}_{2})A_l(\boldsymbol{k}_{3})A_m(\boldsymbol{k}_{4})\rangle .
\end{align}
Thus, the magnetic field four-point correlator can be factorised into a product of projection tensors acting on the underlying gauge field correlator. For notational convenience, we denote the contraction of two magnetic fields in momentum space as a dot product, i.e., $B_\mu(\boldsymbol{k}_1) B^\mu(\boldsymbol{k}_2) 
= \boldsymbol{B}(\boldsymbol{k}_1)\cdot \boldsymbol{B}(\boldsymbol{k}_2)
$. With this, we can define the corresponding trispectrum as follows
\beq
\label{b-tripsectrum}
\langle \boldsymbol{B}(\boldsymbol{k}_1)\cdot \boldsymbol{B}(\boldsymbol{k}_2)\;\boldsymbol{B}(\boldsymbol{k}_3)\cdot \boldsymbol{B}(\boldsymbol{k}_4)\rangle &= \l 2\pi\r^{3}\delta^{(3)}\l \sum_{\alpha=1}^{4}\bk_{\alpha}\r \mathcal{T}_{\zeta B}(\boldsymbol{k}_1,\boldsymbol{k}_2,\boldsymbol{k}_3,\boldsymbol{k}_4)\,,
\eeq
where the full trispectrum $\mathcal{T}_{\zeta B}$ must account for all the permutations, as listed in eq.~(\ref{autocorr}).
For this, we need to identify the contribution from various combinations of the external legs. 
Borrowing the terminology from the flat-space calculations, we can refer to them as distinct channels. Thus, the trispectrum will have contributions from the first channel $[1]:~(12)(34)$, the second channel  $[2]:~(13)(24)$, and the third channel $[3]:~(14)(23)$, respectively. 
The final trispectrum can be written as
\beq\label{bautocorr}
 \mathcal{T}_{\zeta B}&=\frac{1}{a^8}\l\frac{\dot{\lambda}}{H \lambda}\r^2\sum_{S \in\{1,2,3\}} \sum_{X,Y \in \{A,A’\}} Q^{\zeta}_{SXY}\;I^{SXY},
\eeq
where $S=\{1,2,3\}$ stands for the first, second and third channels, respectively. The integrals remain the same as in eq. (\ref{autocorr}). However, we have introduced the superscript $S$ to indicate their respective channel. If we replace $k_2 \leftrightarrow k_3$ in $I^{1XY}$, we would obtain $I^{2XY}$. Similarly, replacing $k_2 \leftrightarrow k_4$  would lead to $I^{3XY}$. Thus, we just need to evaluate the integral in one channel while the others can simply be obtained by these interchanges. For this reason, we emphasise upon the evaluation of the trispectrum in the first channel. The coefficients $Q^\zeta$ are obtained by performing the contraction with generalized projection matrices. For the first channel, they simplify to take the following form,
\begin{align}
\label{QB2}
Q^{\zeta}_{1 AA}&=  k_{1}^{2} k_{2}^{2} k_{3}^{2} k_{4}^{2} 
\l 1 + \cos^{2}\theta_{12} \r 
\l 1 + \cos^{2} \theta_{34} \r~,\\
\label{QB3}
Q^{\zeta}_{1 AA'}&= 2 k_{1}^{2} k_{2}^{2} k_{3} k_{4}\, \l 1 + \cos^{2} \theta_{12}\r\; \cos \theta_{34}~,\\
\label{QB4}
Q^{\zeta}_{1 A'A}&= 2 k_{1} k_{2} k_{3}^{2} k_{4}^{2} \,
\cos\theta_{12} 
\l 1 + \cos^{2}\theta_{34}  \r~,\\
\label{QB1}
Q^{\zeta}_{1 A'A'}&= 4 k_{1} k_{2} k_{3} k_{4} \,\cos\theta_{12} \cos \theta_{34} ~.
\end{align}
In these expressions, $k_1$, $k_2$, $k_3$, and $k_4$ represent the magnitudes of the momenta associated with the magnetic field. Since momentum conservation dictates that all vectors must sum to zero, the momenta collectively form a closed quadrilateral, as shown in Fig.~\ref{quadrilateral}.
Note that, unlike in the case of the three-point function where momentum conservation confines the configuration to a plane, the quadrilateral formed by the four external momenta in the trispectrum is not necessarily planar.
The angles $\theta_{12}$ and $\theta_{34}$ correspond to the angles between the vectors $\boldsymbol{k}_1$ and $\boldsymbol{k}_2$, and $\boldsymbol{k}_3$ and $\boldsymbol{k}_4$, respectively.  One can obtain the coefficients in other channels which are listed in  Appendix \ref{Polcoeffs}. 
For calculational convenience, we define the exchange momenta in each channel with appropriate subscripts, i.e., $\boldsymbol{k}_{\mathrm{I}}=\boldsymbol{k}_1+\boldsymbol{k}_2$, $\boldsymbol{k}_{\mathrm{II}}=\boldsymbol{k}_1+\boldsymbol{k}_3$, and $\boldsymbol{k}_{\mathrm{III}}=\boldsymbol{k}_1+\boldsymbol{k}_4$. These will be used to denote the exchange momenta in the corresponding channels.

Now, the evaluation of the correlator reduces only to the evaluation of the integrals $I^{SAA’}, I^{SA’A},$ and $I^{SA’A’}$, and the evaluation method is identical for all of them. Consequently, we present the analysis of $I^{1AA}$, but not exclusively to it, as the remaining integrals follow a similar approach and easily translate to other channels. 
From eqs. (\ref{IAAdef})-(\ref{IAA'def}), we notice that the evaluation of $I^{1AA}$ is essentially the evaluation of $I^{1AA}_{++}$ and  $I^{1AA}_{-+}$. 
Using eq. (\ref{Ipp1}), we can write them in terms of mode functions as 
\beq
I^{1AA}_{++}= -&A_{k_{1}}\l\eta_{0}\r A_{k_{2}}\l \eta_{0}\r A_{k_{3}}\l\eta_{0}\r A_{k_{4}}\l\eta_{0}\r \int_{-\infty}^{\eta_{0}}d\eta_{1}\;\lambda\l \eta_{1}\r\, A^{*}_{k_{1}}\l \eta_{1}\r A^{*}_{k_{2}}\l \eta_{1}\r\zeta_{k_{\mathrm{I}}}\l\eta_{1}\r\\
&\times\int_{-\infty}^{\eta_{1}}d\eta_{2}\;\lambda\l \eta_{2}\r\, A^{*}_{k_{3}}\l \eta_{2}\r A^{*}_{k_{4}}\l \eta_{2}\r\zeta^{*}_{k_{\mathrm{I}}}\l\eta_{2}\r + \l k_{1},k_{2}\r\leftrightarrow \l k_{3},k_{4}\r,
\eeq
where we have implemented the time ordering in $G^{\zeta}_{k_{\mathrm{I}}}$ which leads to a nested time integral and $\l k_{1},k_{2}\r\leftrightarrow \l k_{3},k_{4}\r$ term. Similarly, from eq. (\ref{Imp1}), the other integral  becomes
\beq
I^{1AA}_{-+}= &A^*_{k_{1}}\l\eta_{0}\r A^*_{k_{2}}\l \eta_{0}\r A_{k_{3}}\l\eta_{0}\r A_{k_{4}}\l\eta_{0}\r \int_{-\infty}^{\eta_{0}}d\eta_{1}\;\lambda\l \eta_{1}\r\, A_{k_{1}}\l \eta_{1}\r A_{k_{2}}\l \eta_{1}\r\zeta_{k_{\mathrm{I}}}\l\eta_{1}\r\\
&\times\int_{-\infty}^{\eta_{0}}d\eta_{2}\;\lambda\l \eta_{2}\r\, A^{*}_{k_{3}}\l \eta_{2}\r A^{*}_{k_{4}}\l \eta_{2}\r\zeta^{*}_{k_{\mathrm{I}}}\l\eta_{2}\r .
\eeq
We are interested in correlators that are evaluated at the time surface $\eta_{0}\rightarrow 0$. In this limit, the mode function with time argument $\eta_{0}$ can be written as,
$A_{k_{1}}\l0\r A_{k_{2}}\l 0\r A_{k_{3}}\l0\r A_{k_{4}}\l0\r=A^*_{k_{1}}\l0\r A^*_{k_{2}}\l0\r A_{k_{3}}\l0\r A_{k_{4}}\l0\r \sim |A_{k_1}^{(0)}||A_{k_2}^{(0)}||A_{k_3}^{(0)}||A_{k_4}^{(0)}|$.
In other words, the mode function outside the integral can be written in terms of the absolute value of the asymptotic limit of the mode function when we evaluate the correlator at the end of inflation, $\eta_{0}\rightarrow 0$. This leads to
\begin{eqnarray}
\label{IA}
I^{1AA}=\frac{\pi^{2}}{8}\;|\zeta_{k_I}^{(0)}|^{2}|A_{k_1}^{(0)}||A_{k_2}^{(0)}||A_{k_3}^{(0)}||A_{k_4}^{(0)}|\;  \tilde{I}^{1AA}~,
\end{eqnarray}
with $\tilde{I}^{1AA}=\text{Re}\ll\tilde{I}^{1AA}_{-+}-\tilde{I}^{1AA}_{++}\rr$, 
\begin{align}
\label{TildeAApp}
\tilde{I}^{1AA}_{++}&= \int_{-\infty}^{\eta_{0}}d\eta_{1}\; \eta_{1}\l 1+ik_{\mathrm{I}}\eta_{1}\r e^{-ik_{\mathrm{I}}\eta_{1}}H_{n+\frac{1}{2}}^{(2)}\l -k_{1}\eta_{1}\r H_{n+\frac{1}{2}}^{(2)}\l- k_{2}\eta_{1}\r \nn\\
\!\!\!\!\!\!&\times \!\!\!\int^{\eta_{1}}_{-\infty}d\eta_{2}\; \eta_{2}\l 1-ik_{\mathrm{I}}\eta_{2}\r e^{ik_{\mathrm{I}}\eta_{2}}H_{n+\frac{1}{2}}^{(2)}\l -k_{3}\eta_{2}\r H_{n+\frac{1}{2}}^{(2)}\l -k_{4}\eta_{2}\r +\l k_{1},k_{2}\r\!\leftrightarrow \l k_{3},k_{4}\r \\
\tilde{I}^{1AA}_{-+}&= \int^{\eta_{0}}_{-\infty}d\eta_{1}\; \eta_{1}\l 1+ik_{\mathrm{I}}\eta_{1}\r e^{-ik_{\mathrm{I}}\eta_{1}}H_{n+\frac{1}{2}}^{(1)}\l -k_{1}\eta_{1}\r H_{n+\frac{1}{2}}^{(1)}\l -k_{2}\eta_{1}\r \nn\\
&\times \int^{\eta_{1}}_{-\infty}d\eta_{2}\; \eta_{2}\l 1-ik_{\mathrm{I}}\eta_{2}\r e^{ik_{\mathrm{I}}\eta_{2}}H_{n+\frac{1}{2}}^{(2)}\l - k_{3}\eta_{2}\r H_{n+\frac{1}{2}}^{(2)}\l -k_{4}\eta_{2}\r.
\label{TildeAAmp}
\end{align}
Similarly, recasting $I^{AA’}, I^{A’A}$ and $I^{A’A’}$ also in terms of mode function asymptotics yields the following structure
\begin{eqnarray}
\label{allIA}
I^{1AA'}&=& \frac{\pi^{2}}{8}\;|\zeta_{k_I}^{(0)}|^2|A_{k_1}^{(0)}||A_{k_2}^{(0)}||A_{k_3}^{(0)}||A_{k_4}^{(0)}|\; k_{3}k_{4}\; \tilde{I}^{1AA'},\\
I^{1A'A}&=& \frac{\pi^{2}}{8}\;|\zeta_{k_I}^{(0)}|^2|A_{k_1}^{(0)}||A_{k_2}^{(0)}||A_{k_3}^{(0)}||A_{k_4}^{(0)}|\; k_{1}k_{2}\; \tilde{I}^{1A'A} ,\\
I^{1A'A'}&=&\frac{\pi^{2}}{8}\;|\zeta_{k_I}^{(0)}|^2|A_{k_1}^{(0)}||A_{k_2}^{(0)}||A_{k_3}^{(0)}||A_{k_4}^{(0)}| \;k_{1}k_{2}k_{3}k_{4}\; \tilde{I}^{1A'A'}.
\end{eqnarray}
Similarly to the earlier case, the reduced integrals $\tilde{I}$'s are, 
\begin{equation}
\tilde{I}^{1AA'}=\text{Re}\ll\tilde{I}^{1AA'}_{-+}-\tilde{I}^{1AA'}_{++}\rr, \;\;\tilde{I}^{1A'A}=\text{Re}\ll\tilde{I}^{1A'A}_{-+}-\tilde{I}^{1A'A}_{++}\rr,\;\;
\tilde{I}^{1A'A'}=\text{Re}\ll\tilde{I}^{1A'A'}_{-+}-\tilde{I}^{1A'A'}_{++}\rr. 
\end{equation}
with, 
\begin{eqnarray}
\label{TildeAA'pp}
\tilde{I}^{1AA'}_{++}&=& \int^{\eta_{0}}_{-\infty}d\eta_{1}\; \eta_{1}\l 1+ik_{\mathrm{I}}\eta_{1}\r e^{-ik_{\mathrm{I}}\eta_{1}}H_{n+\frac{1}{2}}^{(2)}\l -k_{1}\eta_{1}\r H_{n+\frac{1}{2}}^{(2)}\l -k_{2}\eta_{1}\r \\
\!\!\!\!\!\!&\times& \!\!\!\int^{\eta_{1}}_{-\infty}d\eta_{2}\; \eta_{2}\l 1-ik_{\mathrm{I}}\eta_{2}\r e^{ik_{\mathrm{I}}\eta_{2}}H_{n-\frac{1}{2}}^{(2)}\l -k_{3}\eta_{2}\r H_{n-\frac{1}{2}}^{(2)}\l -k_{4}\eta_{2}\r +\l k_{1},k_{2}\r\!\leftrightarrow \l k_{3},k_{4}\r ,\nn\\
\tilde{I}^{1AA'}_{-+}&=&  \int^{\eta_{0}}_{-\infty}d\eta_{1}\; \eta_{1}\l 1+ik_{\mathrm{I}}\eta_{1}\r e^{-ik_{\mathrm{I}}\eta_{1}}H_{n+\frac{1}{2}}^{(1)}\l -k_{1}\eta_{1}\r H_{n+\frac{1}{2}}^{(1)}\l -k_{2}\eta_{1}\r \nn\\
&\times &\int^{\eta_{1}}_{-\infty}d\eta_{2}\; \eta_{2}\l 1-ik_{\mathrm{I}}\eta_{2}\r e^{ik_{\mathrm{I}}\eta_{2}}H_{n-\frac{1}{2}}^{(2)}\l - k_{3}\eta_{2}\r H_{n-\frac{1}{2}}^{(2)}\l -k_{4}\eta_{2}\r.
\label{TildeAA'mp}
\end{eqnarray}
Note that $\tilde{I}^{1A'A}_{++}$ and $\tilde{I}^{1A'A}_{-+}$ are obtained by interchanging $\l k_{1},k_{2}\r \leftrightarrow \l k_{3},k_{4}\r$ in the above expressions. Similarly, the introduced axiliary integrals $\tilde{I}^{A'A'}_{++}$ and $\tilde{I}^{A'A'}_{-+}$ can be obtained by replacing $H_{n+\frac{1}{2}}^{(1,2)}(x)$ with $H_{n-\frac{1}{2}}^{(1,2)}(x)$ in eqs. (\ref{TildeAApp})-(\ref{TildeAAmp}) in an analogous manner. One could implement the evaluation of these integrals analytically or in Mathematica. As we choose the power-law index of the coupling function to be an integer $n$, only Hankel functions of half-integer order appear. For these, one can use an exact closed-form (finite) power-law expansion as \footnote{For non-integer $n$, eq. \ref{hankel_int} is an asymptotic series. However, when $n$ is an integer, the expression becomes exact and reduces to a finite sum, see eq. 10.47.6 and eq. 10.49.7 of \cite{NIST:DLMF} or eq. 14.151 and eq. 14.162 of \cite{ARFKEN2013643}}
\beq\label{hankel_int}
H_{n+\frac{1}{2}}^{(2)}\l z\r \= \l i\r^{n+1}\l \frac{2}{\pi z}\r^{\frac{1}{2}}e^{-iz}\sum_{m=0}^{n}\frac{c_{m}}{z^{m}}\quad \text{with}\quad c_{m} = \frac{(-i)^{m}(n+m)!}{2^{m}m!(n-m)!}.
\eeq
This is because we are only interested in the half-integer (i.e., integer values of $n$) case as we  have chosen the coupling function with its power law index being an integer. 
We list these integral results for different values of $n$ in Appendix~\ref{intresults}.

To summarise, our final result for the magnetic field trispectrum is expressed in eqs.~(\ref{b-tripsectrum}) and  (\ref{bautocorr}) which can now be readily evaluated for various phenomenologically interesting values of $n$.
For instance, $n=2$ yields a scale-invariant magnetic spectral energy density in the case of inflationary magnetogenesis and thus, serves as an interesting example. After evaluating the integrals for a given value of $n$, we can write the complete result for the trispectrum. In a later section, we shall discuss some interesting momentum configurations of the trispectrum for different values of $n$.

For completeness, using our formalism, we can also evaluate the four-point autocorrelation function for electric fields. 
In terms of the auxiliary integrals defined earlier, the full result for the trispectrum can be expressed as follows 
\begin{eqnarray}
\label{eautocorr}
    \langle \boldsymbol{E}(\boldsymbol{k}_1)\cdot \boldsymbol{E}(\boldsymbol{k}_2)\;\boldsymbol{E}(\boldsymbol{k}_3)\cdot \boldsymbol{E}(\boldsymbol{k}_4)\rangle &=& \l 2\pi\r^{3}\delta^{(3)}\l \sum_{\alpha=1}^4\boldsymbol{k}_\alpha\r \l\frac{\dot{\lambda}}{H \lambda}\r^2\frac{1}{a^8}|\zeta_{k_I}^{(0)}|^2|A_{k_1}^{\prime(0)}||A_{k_2}^{\prime(0)}||A_{k_3}^{\prime(0)}||A_{k_4}^{\prime(0)}|\nn\\&\times&k_{1} k_{2} k_{3} k_{4} \l\frac{\pi^{2}}{8} \sum_{S} \sum_{X,Y} \tilde{Q}^{\zeta}_{SXY}\;\tilde{I}^{SXY}\r.
\end{eqnarray}
The corresponding coefficients $\tilde{Q}$'s for the first channel are  
\begin{align}
\label{QE2}
\tilde{Q}^{\zeta}_{1 AA}&= 4\cos\theta_{12} \cos \theta_{34}~,\\
\label{QE3}
\tilde{Q}^{\zeta}_{1 AA'}&= 2 \,
\cos\theta_{12} 
\l 1 + \cos^{2}\theta_{34}  \r~, \\
\label{QE4}
\tilde{Q}^{\zeta}_{1 A'A}&= 2  \,\l 1 + \cos^{2} \theta_{12}\r\, \cos \theta_{34}~,\\
\label{QE1}
\tilde{Q}^{\zeta}_{1 A'A'}&=   \,\l 1 + \cos^{2}\theta_{12} \r 
\l 1 + \cos^{2} \theta_{34} \r~.
\end{align}
The remaining coefficients in the respective channels are listed in the Appendix \ref{Polcoeffs}.
It is important to note that, for our choice of the coupling function, the electric field energy density decays rapidly for positive values of $n$ during inflation \cite{Ferreira:2013sqa, Ferreira:2014hma}. As a result, the electric field trispectrum will be much suppressed as compared to the magnetic field trispectrum, and therefore, in the rest of the paper, we shall not discuss further about it and only make some remarks about its relevance in certain momentum configurations.
\subsubsection{Exploring specific shapes of the magnetic trispectrum}
\label{specific-shapes}
To extract the physical content of the trispectrum, it is often useful to evaluate it in specific momentum configurations that highlight distinct features associated with the underlying interactions. 
In this section, we shall discuss two of the most interesting cases of the trispectrum which are the \textit{equisided configuration} and the \textit{counter collinear} limit. 

\subsubsection*{Equisided configuration}
The equisided configuration refers to a momentum quadrilateral where all external momenta have equal magnitudes, i.e.,
$|\boldsymbol{k}_1|=|\boldsymbol{k}_2|=|\boldsymbol{k}_3|=|\boldsymbol{k}_4|$. This configuration is particularly useful because it maximizes the signal for interactions that involve higher derivative operators or particle exchanges during inflation. The symmetric setup removes directional hierarchies and allows one to isolate the intrinsic shape dependence of the trispectrum without being dominated by kinematic asymmetries. It serves as the natural analogue of the equilateral triangle in the bispectrum, and is particularly useful for comparing the amplitude of different interaction terms in a model-independent manner. 
In other words, the trispectrum can be analysed in terms of two characteristic scales: the equal magnitude of the gauge field momenta and the momentum carried by the exchanged particle.
\begin{figure}[t]
\vskip -20pt
\centering
\includegraphics[scale=0.3]{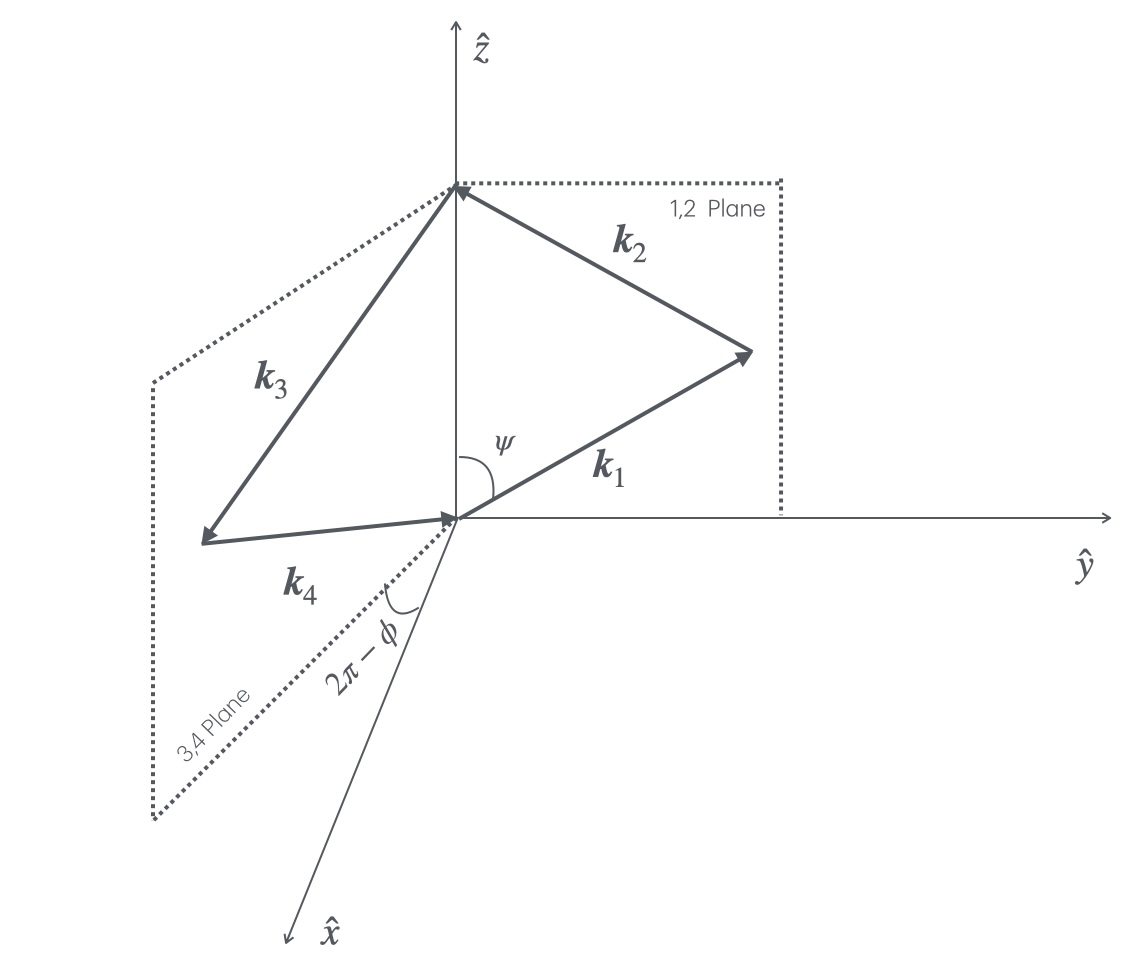}
\caption{A general orientation of the momentum quadrilateral associated with the trispectrum. The momenta $\boldsymbol{k}_1$ and $\boldsymbol{k}_2$ lie in one plane (1,2 plane), while $\boldsymbol{k}_3$ and $\boldsymbol{k}_4$ lie in another plane (3,4 plane).}
\label{coordinate.001}
\end{figure}

In order to fully evaluate the trispectrum in the equisided configuration, we need to specify the behavior of the kinetic coupling function. Since $n=2$ yields a scale-invariant spectral energy density for magnetic fields, we analyse the four-point equilateral correlator for this case which can be easily extended to any other value of $n$ as well. 
For this case, we first need to evaluate the integrals corresponding to the correlator which have been discussed in the previous section. The integrals have been evaluated in an arbitrary kite shaped configuration for different values of $n$, and the corresponding results have been presented in the Appendix~\ref{intresults}. From these results, one can easily take the equisided limit, where the magnitudes of all external momenta are equal, i.e., $k_{1}=k_{2}=k_{3}=k_{4} = k$.

Moreover, to evaluate the trispectrum, one also has to specify the relative angles between four momenta, i.e., $C_{ij}=\cos\theta_{ij}$. For this reason, we fix a coordinate system in the momentum space. We choose the internal momentum $\boldsymbol{k}_{I}$ to define the polar $z$ axis, and define $y\!-\!z$ plane as the plane of $ \boldsymbol{k}_{1}$ and $\boldsymbol{k}_{2}$. The remaining pair, $\boldsymbol{k}_{3}$ and $\boldsymbol{k}_{4}$, is defined to lie in a second plane obtained by rotating the $y\!-\!z$ plane about the $z$-axis by an azimuthal angle $\phi$. Equivalently, the line where this plane intersects the $x\!-\!y$ plane makes an angle $\phi$ with the $x$-axis (see Fig.~\ref{coordinate.001}). It can be seen that all relative angles can be expressed in terms of the polar angle $\psi$ of $\bk_1$ and azimuthal angle $\phi$ as
\begin{align}
    C_{12}&=C_{34}=\cos 2\psi~,\\
      C_{13}&=C_{24}=\sin^2\psi \sin \phi-\cos^2\psi~,\\
      C_{14}&=C_{23}=-\sin^2\psi \sin \phi-\cos^2\psi~.
\end{align}
The cosine of angle $\psi$ is the ratio between the momenta $k$ and $k_{\mathrm{I}}$. Consequently, the total trispectrum in the equilateral shape can be fully specified in terms of the variables $k$, $k_{\mathrm{I}}$, and $\phi$. Depending on the choice of $\phi$, one may obtain the collapsed or counter collinear limit of other channel momenta. Equivalently, the momenta $k_{\mathrm{II}}$ and $k_{\mathrm{III}}$ may vanish for certain values of the azimuthal angle $\phi$. Using above relations, we observe that in the equisided case
\begin{eqnarray}
    k_\mathrm{I}=k \,\sqrt{2\l1+C_{12}\r},\;\;k_\mathrm{II}=k \sqrt{2\l1+C_{13}\r}, \;\;k_\mathrm{III}&=&k \sqrt{2\l1+C_{14}\r}~.
\end{eqnarray}
\begin{figure}[t]
\centering
\includegraphics[width=0.7\linewidth]{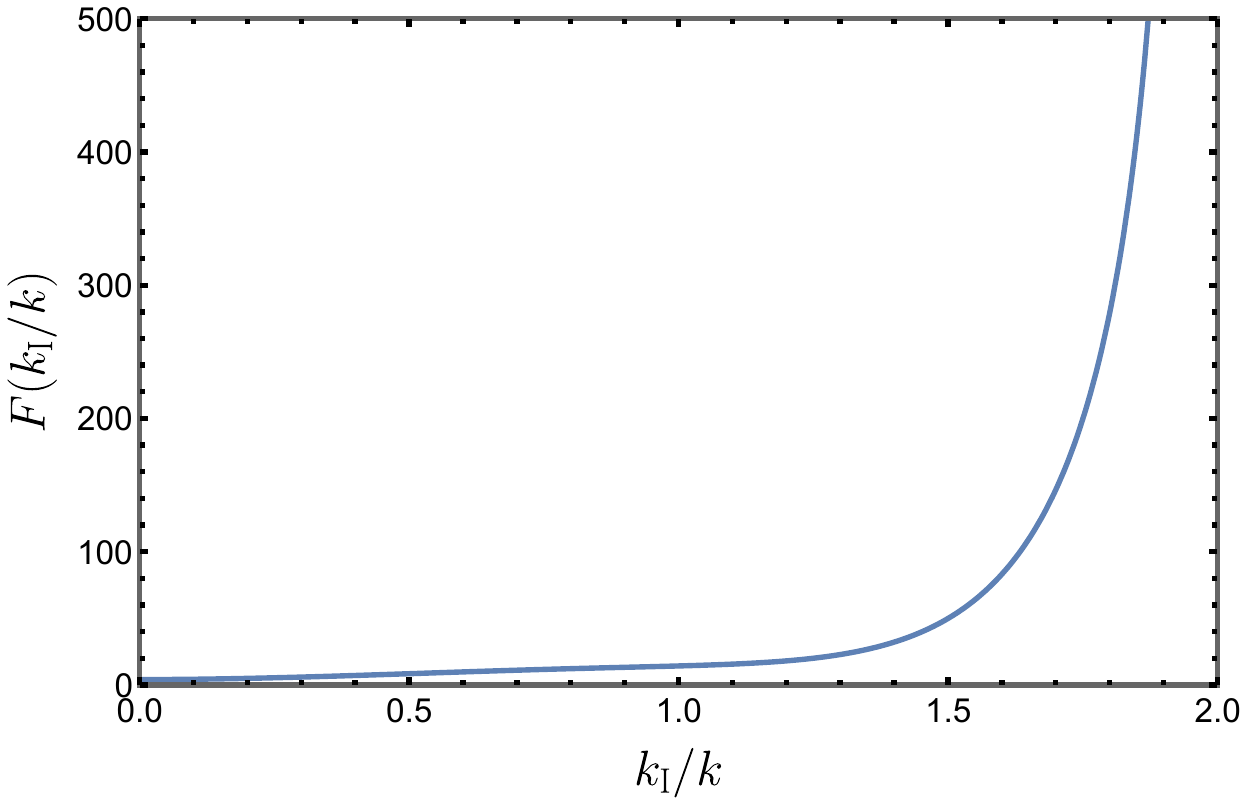}
\caption{The signal strength $F(k_{\mathrm I}/k)$ for $\phi=0$ associated with the equisided configuration is plotted as a function of $k_{\mathrm I}/k$. We found that the signal increases monotonically as the momentum quadrilateral approaches the flattened shape from the counter collinear configuration, and reaches its maximum value when the exchange momentum is at its maximum.}
\label{fig:placeholder}
\end{figure}
For example, $\phi=\pi/2$ results in $k_{\mathrm{III}} \to 0$, leading to the dominance of the third channel over the others. In the subsequent part of this section, we will discuss the detailed analysis of this limit in the first channel exchange momenta. To achieve this, we work with a particular simple choice of $\phi$, ensuring that none of the internal momenta vanish, and analyse the resulting trispectrum. Finally, we choose the momentum quadrilateral in such a way that the two planes containing $\bk_1,\bk_2$ and $\bk_3,\bk_4$ are orthogonal, i.e., $\phi=0$. This ensures that all the internal momenta are finite. With this convention, we can express the trispectrum as
\begin{align}
\langle \boldsymbol{B}(\boldsymbol{k}_1)\cdot \boldsymbol{B}(\boldsymbol{k}_2)\;\boldsymbol{B}(\boldsymbol{k}_3)\cdot \boldsymbol{B}(\boldsymbol{k}_4)\rangle &= \l 2\pi\r^{3}\delta^{(3)}\l \sum_{\alpha=1}^{4}\bk_{\alpha}\r\,\l\frac{\dot{\lambda}}{H \lambda}\r^2 P_\zeta (k_I)P^2_B(k)\; F\left(\frac{k_I}{k},\phi\right)~,
\end{align}
where the net trispectrum amplitude from all channels is given by
\begin{eqnarray}
  \frac{1}{k_{\mathrm{I}}^3} F(k,k_{\mathrm{I}},\phi)=\frac{1}{k_{\mathrm{I}}^3}\mathcal{F}_1(k,k_{\mathrm{I}},\phi)+\frac{1}{k_{\mathrm{II}}^3}\mathcal{F}_2(k,k_{\mathrm{I}},\phi)+ \frac{1}{k_{\mathrm{III}}^3}\mathcal{F}_3(k,k_{\mathrm{I}},\phi)~,
\end{eqnarray}
with
\begin{align}
\mathcal{F}_S(k,k_{\mathrm{I}},\phi) = \frac{\pi^{2}}{32\,k^4}\;\l Q^{\zeta}_{SAA}\;\tilde{I}^{AA}+k^4Q^{\zeta}_{SA'A'}\;\tilde{I}^{A'A'}+k^2Q^{\zeta}_{SAA'}\;\tilde{I}^{AA'}+k^2Q^{\zeta}_{SA'A}\;\tilde{I}^{A'A}\r~.
\end{align}
As a consequence of the triangle inequality, the ratio $k_I/k$ is bounded from above, i.e., $k_I\leq k_1+k_2=2k$ thus, the maximum value of $k_I$ in the equisided case is $k_I=2k$. In Fig.~\ref{fig:placeholder}, we plot the strength of the trispectrum signal characterised by $F(k,k_I,\phi)$ for $\phi=0$ and find that this signal increases monotonically with $k_I/k$ as the momentum quadrilateral approaches the flattened shape from the counter collinear configuration. We further observe that the signal is minimum when the exchange momentum is smallest (counter collinear limit) while it reaches its maximum value in the flattened shape. 
These results can be easily extended to the case of electric field trispectrum as well as for the tensor exchange contribution. However, as we shall argue later,  we expect these contributions to be subdominant for the growing kinetic coupling and due to the smallness of the tensor-to-scalar ratio.  

\subsubsection*{Counter collinear limit}
The counter collinear limit, on the contrary, probes the situation in which two of the external momenta nearly cancel each other, for instance $\boldsymbol{k}_1 \approx -\boldsymbol{k}_2$, so that the exchanged momentum $\boldsymbol{k}_{\mathrm{I}}=\boldsymbol{k}_1+\boldsymbol{k}_2$ becomes soft, i.e., $k_I\to 0$. This configuration captures the long-wavelength modulation of short-wavelength modes, making it especially sensitive to exchange diagrams and multi-field dynamics. In terms of a non-linearity parameter $\beta_{\rm NL}$, the magnetic field trispectrum in this limit can be defined as 
\begin{eqnarray}
\!\!\!\!\!\!\! \lim_{k_{\mathrm{I}}\to0}   \langle \boldsymbol{B}(\boldsymbol{k}_1)\cdot \boldsymbol{B}(\boldsymbol{k}_2)\;\boldsymbol{B}(\boldsymbol{k}_3)\cdot \boldsymbol{B}(\boldsymbol{k}_4)\rangle =\l 2\pi\r^{3}\delta^{(3)}\l\sum_{\alpha=1}^{4}\bk_{\alpha}\r \beta_{NL} P_\zeta (k_I)P_B(k_1)P_B(k_3).
\end{eqnarray}
The counter collinear limit, therefore, is an explicit probe of the contribution of exchange diagrams and carries direct physical meaning, embodying the soft theorems or consistency relations which connect the trispectrum to lower-order correlators.
Unlike the equisided configuration, the four-point correlation in the counter collinear limit can be easily derived for a generic coupling by repeated use of the equation of motion of the gauge fields, similar to the squeezed limit evaluation of the cross-correlation bispectra~\cite{Jain:2012ga, Jain:2012vm, Jain:2021pve, Sai:2023vyf}. 

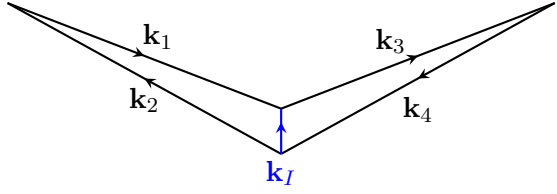
\begin{figure}[t]
\centering
\begin{tikzpicture}[scale=4.0, thick, >=latex]
\coordinate (A) at (0,0);           
\coordinate (B) at (-0.9,.5);        
\coordinate (C) at (0,0.15);       
\coordinate (D) at (.9,.5);       
\draw[black,->-=0.5,>=stealth] (A) -- (B) node[pos=0.45,above=10pt] {$\vb{k}_1$};
\draw[black,->-=0.5,>=stealth] (B) -- (C) node[pos=0.5, below=8pt] {$\vb{k}_2$};
\draw[black,->-=0.5,>=stealth] (C) -- (D) node[pos=0.4, above=1pt] {$\vb{k}_3$};
\draw[black,->-=0.5,>=stealth] (D) -- (A) node[pos=0.5, below=3pt] {$\vb{k}_4$};
\draw[blue,->-=0.7,>=stealth] (A) -- (C) node[midway, below=7pt] {\textcolor{blue}{$\vb{k}_I$}};
\end{tikzpicture}
\caption{The counter collinear limit ($k_I \to 0$) of the momentum quadrilateral associated with the trispectrum. As mentioned earlier, the two triangles need not be coplanar.}
\label{counter-collinear}
\end{figure}

In what follows, we focus on the magnetic field autocorrelation sourced by scalar exchange. Our approach, however, is completely general and applies equally to the electric or gauge field autocorrelation functions. Since we are free to choose any exchange momenta to become soft, without loss of generality, we take $\boldsymbol{k}_{\mathrm{I}}$ to be soft. 
Among the possible exchange channels, we find that the first channel contribution dominates over the others. This dominance arises because, in the counter collinear limit, the kinematic configuration enhances the first channel, while the other two are suppressed. Consequently, it suffices to evaluate only this leading channel for the final result. 
We start by rewriting the polarisation coefficients $Q^\zeta$s of eqs. (\ref{QB2})-(\ref{QB1}) in the counter collinear limit as
\begin{eqnarray}
Q^{\zeta}_{1 AA}= 4k^4 p^4, \;\;
Q^{\zeta}_{1 AA'}= -4 k^4 p^2,\;\;
Q^{\zeta}_{1 A'A} =- 4 k^2 p^4,\;\; 
Q^{\zeta}_{1 A'A'}= 4 k^2 p^2 ~,
\end{eqnarray}
where we made use of the fact that in the counter collinear limit, as in Fig.~\ref{counter-collinear},  $\boldsymbol{k}_1 \approx -\boldsymbol{k}_2 \Rightarrow k_1=k_2=k$, and $\boldsymbol{k}_3 \approx -\boldsymbol{k}_4 \Rightarrow k_3=k_4=p$, which also leads to $\cos \theta_{12}=\cos \theta_{34}=-1$.
Using these explicit forms of the coefficients, one can rewrite the trispectrum as
\beq
\langle \boldsymbol{B}(\boldsymbol{k}_1)\cdot \boldsymbol{B}(\boldsymbol{k}_2)\;\boldsymbol{B}(\boldsymbol{k}_3)\cdot \boldsymbol{B}(\boldsymbol{k}_4)\rangle &\approx\l 2\pi\r^{3}\delta^{(3)}\l \sum_{\alpha=1}^{4}\bk_{\alpha}\r \l\frac{\dot{\lambda}}{H \lambda}\r^2 \\&\, \times k^{2}p^{2} \l I^{1A'A'}+ k^{2}p^{2}I^{1AA}-k^{2}I^{1AA'}-p^{2}I^{1A'A}\r.
\eeq
Here, the above integrals are defined as $I^{1XY} = 2\,\text{Re}\ll I^{1XY}_{++} + I^{1XY}_{-+}\rr$. We first examine the contributions from $I^{1XY}_{++}$, which can be easily extended to $I^{1XY}_{-+}$ to obtain the final result. Note that, in the limit $k_{\mathrm{I}}\rightarrow 0$, the graviton mode functions become time independent, which makes the nested time integrals separate into the integrals of the gauge field mode functions. These relations again demonstrate the interplay between bulk and boundary terms in the evaluation of the nested time integrals.
Collecting the integral results, we find the following compact identity for the ``$++$'' branch as
\beq
\label{CCI1}
I^{1A'A'}_{++} + k^{2}p^{2}I^{1AA}_{++} - k^{2}I^{1AA'}_{++} - p^{2}I^{1A'A}_{++}
= -\lambda^{2}\,|A_{k}|^{2}|A_{p}|^{2}\,A_{k}A^{*'}_{k}A_{p}A^{*'}_{p}.
\eeq
Physically, this relation demonstrates that a precise linear combination of the various integrals collapses into a localized pure boundary expression, reflecting the underlying constraints imposed by the equations of motion.
By repeating the same steps for the ``$-+$'' branch, we obtain
\beq
\label{CCI2}
I^{1A'A'}_{-+}+k^{2}p^{2}I^{1AA}_{-+}-k^{2}I^{1AA'}_{-+}-p^{2}I^{1A'A}_{-+}
= \lambda^{2}\,|\zeta_{k_{\mathrm{I}}}|^2|A_{k}|^{2}|A_{p}|^{2}\,A^{*}_{k}A_{k}'A_{p}A^{*'}_{p}.
\eeq
The proofs of these relations, eqs. (\ref{CCI1}) and (\ref{CCI2}), are outlined in the appendix \ref{intidentity}. Then the above two equations lead to  
\begin{align}
     I^{1A'A'}+ k^{2}p^{2}I^{1AA}-k^{2}I^{1AA'}-p^{2}I^{1A'A}&=4 \lambda^{2}\,|\zeta_{k_{\mathrm{I}}}|^2|A_{k}|^{2}|A_{p}|^{2}\;\text{Im}\!\left[A_{k}A^{\prime*}_{k}\right]\text{Im}\!\left[A_{p}A^{\prime*}_{p}\right].
\end{align}
The remaining term $\text{Im}\!\left[A_{k}A^{\prime*}_{k}\right]$ is the Wronskian of the equation of motion (\ref{EOM_A}), which is fixed by the Bunch–Davies normalization as $2\,\text{Im}\left[A_{k} (\eta_0)A^{\prime*}_{k}(\eta_0)\right]= 1/ \lambda(\eta_0)$.  It finally leads to
\beq
\boxed{
\langle \boldsymbol{B}(\boldsymbol{k}_1)\cdot \boldsymbol{B}(\boldsymbol{k}_2)\;\boldsymbol{B}(\boldsymbol{k}_3)\cdot \boldsymbol{B}(\boldsymbol{k}_4)\rangle \approx\l 2\pi\r^{3}\delta^{(3)}\l \sum_{\alpha=1}^{4}\bk_{\alpha}\r \l\frac{\dot{\lambda}}{H \lambda}\r^2 P_\zeta (k_I)P_B(k)P_B(p).}
\eeq
This evidently implies that   
\beq
\beta_{NL}^{\zeta}=\l\frac{\dot{\lambda}}{H\lambda}\r^2= \l b_{NL}^\zeta\r^2,
\eeq
where $b_{NL}^\zeta$ is the non-linearity parameter associated with the cross-correlation bispectrum $\langle \zeta (\boldsymbol{k}_1)  \boldsymbol{B}(\boldsymbol{k}_2) \cdot \boldsymbol{B}(\boldsymbol{k}_3)\rangle$ in the limit $\boldsymbol{k}_1 \to 0$, $\boldsymbol{k}_2 \approx -\boldsymbol{k}_3 \Rightarrow k_2=k_3=k$, defined by
\beq
\langle \zeta (\boldsymbol{k}_1)  \boldsymbol{B}(\boldsymbol{k}_2) \cdot \boldsymbol{B}(\boldsymbol{k}_3)\rangle 
\approx\l 2\pi\r^{3}\delta^{(3)}\l \sum_{\alpha=1}^{3}\vb{k}_{\alpha}\r b_{NL}^\zeta P_\zeta (k_1)P_B(k),
\eeq
with $b_{NL}^\zeta = \dot{\lambda}/{H\lambda}$ \cite{Jain:2012ga, Jain:2012vm,Sai:2023vyf}. 
This non-linearity parameter indicates a logarithmic rate of change of the coupling function per unit Hubble time which is naively expected as the bispectrum is entirely induced by the dynamical coupling function. Moreover, the quadratic relation between $\beta_{NL}^{\zeta}$ and $b_{NL}^\zeta$ is a notable consequence of the interaction structure: while the bispectrum arises from a single trilinear vertex, the trispectrum is generated by two such vertices.
This relation represents a distinctive and novel result of our work. Our findings further reveal that the relation obtained between $\beta_{NL}^{\zeta}$ and $b_{NL}^\zeta$ is similar to the well-known Suyama-Yamaguchi relation, connecting the trispectrum and bispectrum associated with $\zeta$ \cite{Suyama:2007bg, Suyama:2020akr}. Moreover, our relation $\beta_{NL}^{\zeta}= \l b_{NL}^\zeta\r^2$ can be considered as a lower bound, however,  
in a generic interaction, this would result in $\beta_{NL} \geq\l b_{NL}^\zeta\r^2$. The presence of graviton exchange clearly validates this inequality and it can be established in general, as shown in \cite{Assassi:2012zq}. In addition, with regard to $\beta_{NL}^\zeta$, the presence of additional gauge field interactions without slow-roll suppression invalidates the equality condition, and the corresponding relation is generically relaxed to an inequality. 
\subsection{Tensor exchanges}
Having established the scalar exchange contribution, we now turn to the case of tensor exchanges. The essential difference here arises from the fact that the exchanged mode is a spin-2 graviton $\gamma_{ij}$, whose propagator and polarisation tensors lead to a richer angular dependence in the resulting correlator. In contrast to scalar mediation, the tensor exchange contribution explicitly exhibits anisotropy with respect to the direction of exchange momenta in each channel, i.e., carries information about the helicity structure of the graviton. In our case, both helical graviton modes contribute equally, as there are no parity-violating terms in the gravitational sector. Let us recall the mode expansion of the graviton, given by
\beq
\label{gamma_ij}
\gamma_{ij}({\bf x},\eta) 
= \int\frac{d^{3}{\bf k}}{(2\pi)^{3}}\sum_{s = \pm 2}
\left[\gamma_{k}(\eta) \, e^{i{\bf k}\cdot {\bf x}} \,  \epsilon ^{s}_{ij}(\hat{\bf k}) \, b_{\bf k}^s + h.c. \right],
\eeq
where $\epsilon_{ij}^{s}$ is the polarisation tensor corresponding to the helicity $s$, with the normalisation $\epsilon_{ij}^{s} \epsilon_{ij}^{*s'}=2\delta_{ss'}$ and the creation and annihilation operators do satisfy the condition, $[b_{\bf k}^s,b^{s' \dagger}_{\bf k'}]=(2\pi)^3 \delta^{(3)}({\bf k}-{\bf k'}) \delta_{s s'} $. The Fourier mode solution and the power spectrum of tensor perturbations have already been discussed in section~\ref{eq:mps}. 

Similar to the case of scalars, we start with the interaction Hamiltonian $H_{\gamma AA}$ obtained in eq. (\ref{H_gAA}), which couples two gauge fields to a graviton and leads to a similar interaction vertex as for the curvature perturbation, as in Fig.~\ref{vertex} but wavy leg indicating the graviton. Using this in the in-in master formula eq. (\ref{exp1_in-in}), the leading order connected contribution to the four-point autocorrelation function leads to the same diagrams as in Fig.~{\ref{diagram}}. Our working procedure remains the same as in the case of scalar exchanges, with scalar propagators now replaced by tensor propagators. Since we have already analysed in detail various integrals appearing in the scalar exchange, we can readily use them to represent the tensor exchange result instead of defining a new set of integrals. One can immediately notice that integrals appearing in the tensor exchange are the same as in eq.~(\ref{autocorr}), but with replacement of $W^{\zeta}_k(\eta_1, \eta_2)$ and  $G^{\zeta}_k(\eta_1, \eta_2)$  with $W^{\gamma}_k(\eta_1, \eta_2)$ and  $G^{\gamma}_k(\eta_1, \eta_2)$, respectively which are defined as
\begin{eqnarray}
    W^{\gamma}_k(\eta_1, \eta_2)&=&\gamma_k(\eta_1) \gamma^{*}_k(\eta_2),\\
 G_k^\gamma( \eta_1, \eta_2) &=& W_k^\gamma(\eta_1, \eta_2) \, \Theta(\eta_1 - \eta_2) + W_k^\gamma(\eta_2, \eta_1) \, \Theta(\eta_2-\eta_1).
\end{eqnarray}
Moreover, it can be related to the scalar case by a factor of the tensor-to-scalar ratio $r$, as $ W^{\gamma}_k(\eta_1,\eta_2)=r\; W^{\zeta}_k(\eta_1, \eta_2)$. With this, we obtain the result for the four-point autocorrelation function of gauge fields with tensor exchange as 
\beq\label{Tautocorr}
\expval{A_{i}\l \bk_{1}\r A_{j}\l \bk_{2}\r A_{l}\l \bk_{3}\r A_{m}\l \bk_{4}\r}\= r\;\l 2\pi\r^{3}\delta^{(3)}\l \sum_{\alpha=1}^{4}\bk_{\alpha}\r  \sum_{X,Y \in \{A,A’\}} Q^{\gamma XY}_{ijlm}\, I^{XY}\\
+ & \; (j,\bk_{2})\leftrightarrow (l,\bk_{3})+ (j,\bk_{2})\leftrightarrow (m,\bk_{4}).
\eeq
Since these integrals have already been defined and analysed in the previous section, the only difference here lies in the polarisation coefficients $Q^{\gamma XY}_{ijlm}\left(\bk_1,\bk_2,\bk_3,\bk_4;\bk_{\mathrm{I}}\right)$ are given by
\begin{align}
 Q^{\gamma AA}_{ijlm} &=  \alpha_{abcd}\!\left(\boldsymbol{k}_{1},\boldsymbol{k}_{2}\right)\, \Pi_{ic}\!\left(\hat{\boldsymbol{k}}_{1}\right)\,\Pi_{jd}\!\left(\hat{\boldsymbol{k}}_{2}\right)\,\omega_{aba'b'}\!\left(\hat{\boldsymbol{k}}_{\mathrm{I}}\right)\,\Pi_{lc'}\!\left(\hat{\boldsymbol{k}}_{3}\right)\,
\Pi_{md'}\!\left(\hat{\boldsymbol{k}}_{4}\right)\,\alpha_{a'b'c'd'}\!\left(\boldsymbol{k}_{3},\boldsymbol{k}_{4}\right), \\
Q^{\gamma AA'}_{ijlm} &=  \alpha_{abcd}\!\left(\boldsymbol{k}_{1},\boldsymbol{k}_{2}\right)\, \Pi_{ic}\!\left(\hat{\boldsymbol{k}}_{1}\right)\, \Pi_{jd}\!\left(\hat{\boldsymbol{k}}_{2}\right)\, \omega_{aba'b'}\!\left(\hat{\boldsymbol{k}}_{\mathrm{I}}\right)\,\Pi_{la'}\!\left(\hat{\boldsymbol{k}}_{3}\right)\,\Pi_{mb'}\!\left(\hat{\boldsymbol{k}}_{4}\right), \\
Q^{\gamma A'A}_{ijlm} &= \Pi_{ia}\!\left(\hat{\boldsymbol{k}}_{1}\right)\,\Pi_{jb}\!\left(\hat{\boldsymbol{k}}_{2}\right)\,\omega_{aba'b'}\!\left(\hat{\boldsymbol{k}}_{\mathrm{I}}\right)\, \Pi_{lc'}\!\left(\hat{\boldsymbol{k}}_{3}\right)\,\Pi_{md'}\!\left(\hat{\boldsymbol{k}}_{4}\right)\,\alpha_{a'b'c'd'}\!\left(\boldsymbol{k}_{3},\boldsymbol{k}_{4}\right)\\
Q^{\gamma A' A'}_{ijlm} &= \Pi_{ia}\!\left(\hat{\boldsymbol{k}}_{1}\right)\,\Pi_{jb}\!\left(\hat{\boldsymbol{k}}_{2}\right)\,\omega_{aba'b'}\!\left(\hat{\boldsymbol{k}}_{\mathrm{I}}\right)\, \Pi_{la'}\!\left(\hat{\boldsymbol{k}}_{3}\right)\, \Pi_{mb'}\!\left(\hat{\boldsymbol{k}}_{4}\right)~,
\end{align}
with, 
\beq
\omega_{abcd}\l \hat{\vb{q}}\r \= -\Pi_{ab}\l\hat{\vb{q}} \r\Pi_{cd}\l\hat{\vb{q}} \r+\Pi_{ac}\l\hat{\vb{q}} \r\Pi_{bd}\l\hat{\vb{q}} \r+\Pi_{ad}\l\hat{\vb{q}} \r\Pi_{bc}\l\hat{\vb{q}} \r,\\
\alpha_{abcd}\l \vb{k},\vb{p}\r\=\, \delta_{cd}k_{a}p_{b}+ \l \vb{k}\cdot\vb{p}\r\delta_{ac}\delta_{bd}-2 \delta_{bd}k_{a}p_{c} .
\eeq
The polarisation coefficients  $Q^{\gamma XY}_{ijlm}$ structure given above makes it explicit how tensor
exchange differs qualitatively from the scalar case. In the scalar case, the coefficients
reduce to products of simple projection operators $\Pi_{ij}(\hat{\boldsymbol{k}})$, which
enforce transversality but do not introduce additional angular complexity beyond the scalar
products of external wavevectors. On the contrary, in the tensor case, the presence of the graviton
propagator encoded in $\omega_{aba'b'}(\hat{\boldsymbol{q}})$ and the bilinear tensor
structures $\alpha_{abcd}(\boldsymbol{k}_{\alpha},\boldsymbol{k}_{\beta})$ produce
contractions amongst four distinct directions: the internal graviton momentum
$\hat{\boldsymbol{q}}$ and the external momenta
$\hat{\boldsymbol{k}}_{1}, \hat{\boldsymbol{k}}_{2}, \hat{\boldsymbol{k}}_{3},
\hat{\boldsymbol{k}}_{4}$. In the above equations, $\boldsymbol{q}=\bk_{\mathrm{I}}=\bk_1+\bk_2$ represents the exchange momenta in the initial channel. Analogous to the scalar exchange, $\boldsymbol{q}$ will be $\bk_{\mathrm{II}}$ and $\bk_{\mathrm{III}}$ other corresponding channels. 
As a result, the four-point correlator inherits a non-trivial dependence on the relative angles between these momentum vectors and the configuration of the momentum quadrilateral. For instance, one finds that
\begin{itemize}
\item 
Various terms proportional to $\omega_{aba'b'}(\hat{\boldsymbol{q}})$ involve quadratic combinations of projection operators orthogonal to the graviton momentum, thereby correlating the azimuthal orientation of $\hat{\boldsymbol{q}}$ with the external legs.
\item 
The $\alpha_{abcd}$- structures encode mixed contractions such as $(\hat{\boldsymbol{k}}_{1}\!\cdot\!\hat{\boldsymbol{k}}_{2})$ or
$(\hat{\boldsymbol{k}}_{3}\!\cdot\!\hat{\boldsymbol{k}}_{4})$, but dressed with free indices which couple back to the polarisation tensors. This introduces higher-order angular functions (quadrupolar and higher) into different shapes of the trispectrum.
\item 
In specific configurations, e.g., in the counter collinear limit where
$\boldsymbol{k}_{1}\simeq -\boldsymbol{k}_{2}$ and
$\boldsymbol{k}_{3}\simeq -\boldsymbol{k}_{4}$, the tensor structures enhance correlations
aligned along $\hat{\boldsymbol{k}}_{\mathrm{I}}$, whereas in equilateral quadrilaterals of momenta, the spin-2 nature manifests as oscillatory angular dependence which is otherwise absent in the scalar exchange.
\end{itemize}

From a physical perspective, these polarisation tensors effectively project the external
gauge field fluctuations onto \emph{spin-2 spherical harmonics}, such that the resulting trispectrum
carries an imprint of the helicity-2 nature of gravitons. This associated richer angular structure is,
therefore, a clear discriminator between the two. While scalar exchange yields shapes controlled by the isotropic functions of magnitudes and scalar products, tensor exchange injects \emph{directional
information tied to the graviton polarisation}, producing angular modulations that could, in
principle, be observable in precision measurements of anisotropic non-Gaussianities.

We now turn to calculating the magnetic field correlator from the gauge field correlator which  introduces bilinear projectors $\mathbb{P}_{ij}(\boldsymbol{k}_{2},\boldsymbol{k}_{1})$ and
$\mathbb{P}_{mn}(\boldsymbol{k}_{4},\boldsymbol{k}_{3})$ on the external legs, as already seen in eq.~(\ref{A2B}). 
This projection does not alter the internal graviton structure $\omega_{aba'b'}(\hat{\boldsymbol{q}})$
encoded in the coefficients $Q^{\gamma XY}_{ijmn}$. The magnetic field trispectrum can therefore be
written in the same schematic form as the gauge field one, but with new polarisation weights
\begin{align}
\label{TQ1}
Q^{\gamma}_{1XY} &\equiv
\mathbb{P}_{ij}(\boldsymbol{k}_2,\boldsymbol{k}_1)\,
\mathbb{P}_{lm}(\boldsymbol{k}_4,\boldsymbol{k}_3)\,
Q^{\gamma XY}_{ijlm}\left(\bk_1,\bk_2,\bk_3,\bk_4;\bk_{\mathrm{I}}\right)~, \\
\label{TQ2}
Q^{\gamma}_{2XY}& \equiv \mathbb{P}_{ij}(\boldsymbol{k}_2,\boldsymbol{k}_1)\,
\mathbb{P}_{lm}(\boldsymbol{k}_4,\boldsymbol{k}_3)\;
Q^{\gamma XY}_{iljm}\left(\bk_1,\bk_3,\bk_2,\bk_4;\bk_{\mathrm{II}}\right)~,\\
\label{TQ3}
Q^{\gamma}_{3XY}& \equiv \mathbb{P}_{ij}(\boldsymbol{k}_2,\boldsymbol{k}_1)\;
\mathbb{P}_{lm}(\boldsymbol{k}_4,\boldsymbol{k}_3)\;
Q^{\gamma XY}_{imlj}\left(\bk_1,\bk_4,\bk_3,\bk_2;\bk_{\mathrm{III}}\right)~.
\end{align}
Here, $XY$ super/subscripts denote same as in scalar case with $X$ and $Y$ can take values from the sets $\{A, A’\}$ and subscripts $1,2,3$ denote the respective channels.
From these structures, 
several important observations follow,
\begin{itemize}
 \item The projectors
 $\mathbb{P}_{ij}(\boldsymbol{k}_2,\boldsymbol{k}_1)
 =(\boldsymbol{k}_2\!\cdot\!\boldsymbol{k}_1)\,\delta_{ij}-k_{2i}k_{1j}$
depend only on the \emph{external} directions
$\hat{\boldsymbol{k}}_1,\hat{\boldsymbol{k}}_2$ (and similarly for $3,4$).
They introduce additional angular weights such as
$C_{12}\equiv \hat{\boldsymbol{k}}_1\!\cdot\!\hat{\boldsymbol{k}}_2$ and
$C_{34}\equiv \hat{\boldsymbol{k}}_3\!\cdot\!\hat{\boldsymbol{k}}_4$, but
do not affect the spin--2 dependence tied to
$\hat{\boldsymbol{k}}_{\mathrm{I}}=\widehat{\boldsymbol{k}_1+\boldsymbol{k}_2}$ contained
in $\omega_{aba'b'}(\hat{\boldsymbol{k}}_{\mathrm{I}})$.
Hence, the \emph{richer angular dependence} originating from the tensor exchange is preserved.
\item 
In the counter collinear limit, 
$\boldsymbol{k}_{\mathrm{I}}=\boldsymbol{k}_1+\boldsymbol{k}_2\to 0$ (and similarly for $3,4$), one has
$\boldsymbol{k}_2\simeq -\boldsymbol{k}_1$, which gives
$
\mathbb{P}_{ij}(\boldsymbol{k}_2,\boldsymbol{k}_1)
\simeq -k_1^{2}\,\Pi_{ij}(\hat{\boldsymbol{k}}_1),
$
and first channel dominates the corresponding polarisation coefficients takes the form 
\begin{align}
Q^{\gamma}_{1AA}\;&\longmapsto\;
k_1^{4}k_3^{4}\;
\hat{k}_{1a}\,\hat{k}_{1b}
\,
\omega_{aba'b'}(\hat{\bk}_{\mathrm{I}})\,\hat{k}_{3a'}\,\hat{k}_{3b'} ~,\\
Q^{\gamma}_{1AA'}\;&\longmapsto\;
k_1^{4}k_3^{2}\;
\hat{k}_{1a}\,\hat{k}_{1b}
\,
\omega_{aba'b'}(\hat{\bk}_{\mathrm{I}})\,\hat{k}_{3a'}\,\hat{k}_{3b'} ~,\\
Q^{\gamma}_{1A'A}\;&\longmapsto\;
k_1^{2}k_3^{4}\;
\hat{k}_{1a}\,\hat{k}_{1b}
\,
\omega_{aba'b'}(\hat{\bk}_{\mathrm{I}})\,\hat{k}_{3a'}\,\hat{k}_{3b'} ~,\\
Q^{\gamma}_{1A'A'}\;&\longmapsto\;
k_1^{2}k_3^{2}\;
\hat{k}_{1a}\,\hat{k}_{1b}
\,
\omega_{aba'b'}(\hat{\bk}_{\mathrm{I}})\,\hat{k}_{3a'}\,\hat{k}_{3b'}~,
\end{align}
showing explicitly that the spin--2 projection from $\omega$ fully survives the
$A_i\to B$ mapping.
\item 
Away from the collapsed or counter collinear limit, the
$\mathbb{P}$-tensors inject extra $C_{12},C_{34}$ factors (and powers of $k_a$) which \emph{modulate} the tensor exchange anisotropy but do not isotropize it.
For instance, the magnetic trispectrum contribution from the first channel exhibits angular dependence of the schematic form
\[
\big[\,\text{polynomial in }(C_{12},C_{34})\,\big]\times
\big[\,\text{spin-2 kernel in }\hat{\boldsymbol{k}}_{\mathrm{I}}\,\big].
\]
\end{itemize}
In other words, the angular information tied to the graviton polarisation is fully inherited by the magnetic field four-point correlator. The projectors simply provide additional angular weights from external momenta, but the quadrupolar (and higher) features characteristic of tensor exchange remain intact and can even become more pronounced in special momentum configurations. The final result for the total trispectrum from the graviton exchange can be written as
\beq\label{bgautocorr}
 \mathcal{T}_{\gamma B}&=\,\frac{r}{a^8}\,\sum_{S \in\{1,2,3\}} \sum_{X,Y \in \{A,A’\}} Q^{\gamma}_{SXY}\;I^{SXY},
\eeq
where the polarisation weights $Q^\gamma_{SXY}$ have already been discussed in eqs. (\ref{TQ1})-(\ref{TQ3}).
For brevity, we omit their explicit expressions in terms of the relative angles, as they become rather cumbersome. Furthermore, as noted in the previous section, the electric field trispectrum arising from tensor exchange can be straightforwardly obtained from the gauge field result in eq. (\ref{Tautocorr}) by replacing the external legs with the derivatives of the gauge field. However, we have observed that all these tensor exchange signals are further suppressed by the tensor-to-scalar ratio compared to the scalar exchange scenario. Consequently, we refrain from providing a comprehensive analysis of the tensor exchange contribution, unlike the scalar exchange, which we examined in detail.

While the tensor mediated trispectrum of gauge fields remains subdominant, it still induces distinctive angular dependence arising from the graviton polarisation. This can produce novel characteristic anisotropic signatures in the primordial trispectrum. Future cosmological surveys, which are focused on precise measurements of non-Gaussianity in the CMB, can shed light on these distinctive higher-order imprints involving gravitons and gauge fields. 
\section{Conclusions and discussions}
\label{conclusions}
In this paper, we have calculated, for the first time, the four-point autocorrelation function or the connected trispectrum of gauge field fluctuations which are excited during inflation through their kinetic coupling to the inflaton.
Due to the structure of gauge field interactions, the contact diagrams do not exist and therefore, the resulting contribution to the trispectrum arises \emph {exclusively} through the exchange of scalar or tensor perturbation. 
Using the in-in formalism and cosmological diagrammatic rules, we first derived the full result for the trispectrum of gauge fields for both the scalar and tensor exchange diagrams and used it to obtain the trispectra of the associated magnetic and electric fields. Since electric fields remain subdominant for a growing kinetic coupling function as in our scenario, we found that the electric field trispectrum is subdominant compared to the magnetic field trispectrum. 

Further, we examined the magnetic field trispectrum in two different momentum configurations: equisided and counter collinear. 
For an equisided configuration, we found that the resulting non-Gaussian signal is monotonically increasing as the shape approaches the flattened configuration, indicating its importance for cosmological observations. In the counter collinear limit, we showed that the magnetic field trispectrum follows a novel consistency relation and the amplitude of the trispectrum is equal to the square of the non-linearity parameter associated with the cross-correlation bispectrum of curvature perturbation with magnetic fields. This is an important new result, similar to the Suyama-Yamaguchi relation which indicates a hierarchical relationship amongst the higher-order correlation function of primordial fluctuations in an appropriate soft momentum limit. 

In the case of magnetic field trispectrum sourced by tensor exchange, we observe that the trispectrum inherits a richer angular dependence arising from the non-trivial polarisation coefficients, although its overall amplitude is suppressed by the tensor-to-scalar ratio $r$. Following the current bound $r \lesssim 0.036$ \cite{BICEP:2021xfz, Campeti:2022vom}, we conclude that the tensor exchange trispectrum will remain subdominant compared to the scalar exchange trispectrum. However, the richer angular structure associated with the tensor exchange imprints directional information introduced by the graviton polarisation, producing characteristic angular modulations in the trispectrum. 
This distinctive signature can be sought in upcoming cosmological surveys, in particular through precise measurements of non-Gaussianity in the CMB and large scale structure.

In earlier works \cite{Motta:2012rn,Shiraishi:2012xt,Jain:2012vm,Jain:2021pve}, it has been shown that the cross-correlation bispectrum of metric fluctuations with gauge fields (or magnetic fields) exhibits a logarithmic time dependence as ${\rm log}(k\eta_0)$ in the super Hubble limit for the scale invariant case of the magnetic field with $n=2$. Since the exchange interactions are mediated by the same Hamiltonian, one might naively expect a similar time dependence for the trispectrum as well. However, we observe that for $n=2$, the 4-point autocorrelation function does not exhibit a similar time dependence and it only arises beyond $n=2$. In particular, as shown in Appendix \ref{intresults}, the integral evaluation gives a logarithmic dependence for $n=3$ and we expect this behavior to continue for even larger positive values of $n$. This peculiar result can be associated with the asymptotic behavior of Hankel functions in the super-Hubble limit. Needless to mention, a logarithmic time dependence will also result in a large observable signal for the trispectrum.

Our work opens several directions for future exploration. Since higher-order correlations involving gauge fields bring in very interesting signatures, one can explore the trispectrum in other configurations as well as their full shape function. 
In general, mapping out the trispectrum systematically across different momentum configurations would allow for the identification of robust signatures of gauge field dynamics and facilitate the construction of observationally useful templates. 
Finally, it would be interesting to investigate the gauge field trispectrum in scenarios with parity-violating couplings. Such interactions naturally generate anisotropic, parity-breaking signatures in the observable trispectrum, which could be constrained by future cosmological probes such as CMB and 21-cm surveys. We leave these promising avenues for future work.

From an observational perspective, placing direct constraints on the magnetic field trispectrum is challenging. Current observations of intergalactic magnetic fields primarily constrain the field amplitude and its coherence length, rather than higher-order correlation functions. Moreover, the magnetic field trispectrum is not a standard observable, and any realistic sensitivity in the near future is likely to arise only indirectly.
In particular, future measurements of CMB non-Gaussianities --- including the trispectrum --- could provide an indirect probe. High-precision observations of CMB polarization may improve constraints on the CMB trispectrum, which could then be translated into bounds on the magnetic field bispectrum or trispectrum \cite{Trivedi:2011vt, Trivedi:2013wqa}. In addition, future observations of CMB spectral distortions and their cross-correlations with CMB temperature and polarization anisotropies may offer complementary avenues to constrain higher-order magnetic field correlators \cite{Ganc:2014wia}. A detailed quantitative analysis is beyond the scope of the present work.

\section*{Acknowledgments}
RKJ and SRH would like to acknowledge financial support from the Indo-French Centre for the Promotion of Advanced Research (CEFIPRA) for support of the proposal 6704-4 under the Collaborative Scientific Research Programme. RKJ also acknowledges support from the IISc Research Awards 2024 and SERB, Department of Science and Technology, GoI, through the MATRICS grant~MTR/2022/000821. PJS
thanks Basundhara Ghosh for partial support through the DST Inspire Grant during this project.
\section*{Author contributions note}
All authors contributed equally at every stage of this work.
\appendix
\section{Cosmological diagrammatic rules}
\label{CDrules}
In this appendix, we summarise the diagrammatic rules \cite{Giddings:2010ui} for computing the in-in correlators for the scalar exchange just with the $\zeta A^2$ vertex.
\boe
\1 Draw the final time hypersurface, e.g., the end of inflation epoch \( \eta = \eta_{0} \).
\1 Construct all diagrams relevant to the process, allowing interaction vertices to appear both above and below the final time hypersurface.
\1 Assign propagators as follows: lines that cross or terminate on the final time hypersurface are represented by the Wightman propagator:
\beq
W_k(\eta, \eta') = A_k(\eta) A^{*}_k(\eta') ,           
\eeq
where $A_{k}$'s are the mode functions,  $\eta$ corresponds to the uppermost vertex and $\eta'$ to the lowermost. Lines entirely below the final time surface correspond to the Feynman propagator:
\beq
G_F(k; \eta, \eta') \= W_k(\eta, \eta') \, \theta(\eta - \eta') + W_k(\eta', \eta) \, \theta(\eta'-\eta ) \,.
\eeq
Similarly, lines entirely above the final time surface use the time-reversed Feynman propagator:
\beq
G_F(k; \eta', \eta) = G_F^*(k; \eta, \eta') \,.
\eeq
\1 Vertices located below the final time hypersurface contribute a factor of $-2in\lambda$, while those above contribute $2in\lambda$. Momentum is conserved at each vertex, and an overall momentum-conserving delta function is included.
\1 Each vertex insertion is integrated over conformal time with an appropriate measure. 
\1 If the diagram contains symmetry factors, divide by the appropriate factor. Additionally, diagrams related by reflection across the final time surface are accounted for by taking $2\text{Re}$, as they are complex conjugate to each other.
\eo
These rules remain the same for tensor exchange where one has to replace the vertices below the final time surface contributes a factor of $-i$ and those above contributes $+i$.
\section{Polarisation coefficients of magnetic and electric trispectra for different channels}
\label{Polcoeffs}
Introducing the notation $C_{ij}=\cos\theta_{ij}$, the polarisation coefficients for magnetic and electric
trispectra are listed below.
\subsection*{Magnetic Field}
In the first channel are given by
\begin{align}
Q^{\zeta}_{1 AA}&= k_{1}^{2} k_{2}^{2} k_{3}^{2} k_{4}^{2}\; 
\ll 1 +  C_{12}^2  \rr 
\ll 1 +  C_{34}^2 \rr ,\\
Q^{\zeta}_{1 AA'}&= 2 k_{1}^{2} k_{2}^{2} k_{3} k_{4} \; C_{34}\ll 1 +  C_{12}^2\rr ,\\
Q^{\zeta}_{1 A'A}&= 2 k_{1} k_{2} k_{3}^{2} k_{4}^{2} \;
 C_{12} 
\ll 1 +  C_{34}^2 \rr ,\\
Q^{\zeta}_{1 A'A'}&= 4 k_{1} k_{2} k_{3} k_{4} \; C_{12} C_{34} .
\end{align}
For the second channel, the coefficients take the form
\begin{align}
Q^{\zeta}_{2 AA}&= \l k_{1} k_{2} k_{3} k_{4}\r^2\Bigg[\l C_{12}C_{34}\r^{2}
+ \l C_{14}C_{23}\r^{2}
+ \l C_{13}C_{24}\r^{2}\nonumber \\
&- 2\,C_{13}C_{14}C_{23}C_{24}
- 2\,C_{12}C_{14}C_{23}C_{34}- \,C_{12}C_{13}C_{24}C_{34}\nonumber\\
&+ C_{12}C_{13}C_{23} +C_{13}C_{14}C_{34}
+ C_{23}C_{24}C_{34}
+ C_{12}C_{14}C_{24}
\,\Bigg],
\\
Q^{\zeta}_{2 AA'}&=k_1^2k_3^2k_2k_4 \Bigg[C_{13}^{2} C_{24} - C_{13} C_{14} C_{23} + C_{12} C_{14} + C_{23} C_{34}\Bigg] ,\\
Q^{\zeta}_{2 A'A}&=  k_2^2k_4^2k_1k_3\Bigg[C_{13}C_{24}^{2} - C_{23}C_{14}C_{24}+ C_{12}C_{23} + C_{14}C_{34}\Bigg],\\
Q^{\zeta}_{2 A'A'}&=  k_{1} k_{2} k_{3} k_{4} \Bigg[ C_{12}  C_{34} + C_{13}C_{24}\Bigg].
\end{align}
Finally, for the third channel, the coefficients are
\begin{align}
Q^{\zeta}_{3 AA}&= \l k_{1} k_{2} k_{3} k_{4}\r^2\Bigg[\l C_{23}C_{14}\r^{2}
+ C_{13}^{2}C_{24}^{2}
+ C_{12}^{2}C_{34}^{2} \nonumber \\[4pt]
&-2\,C_{13}C_{23}C_{14}C_{24}
-2\,C_{12}C_{13}C_{24}C_{34}
- C_{12}C_{23}C_{14}C_{34} \nonumber \\[4pt]
&+ C_{12}C_{13}C_{23}
+ C_{12}C_{14}C_{24}
+ C_{13}C_{14}C_{34}
+ C_{23}C_{24}C_{34}\Bigg] ,\\
Q^{\zeta}_{3 AA'}&= k_1^2k_4^2k_2k_3\Bigg[C_{23}C_{14}^{2} - C_{13}C_{14}C_{24} + C_{12}C_{13} + C_{24}C_{34}\Bigg],\\
Q^{\zeta}_{3 A'A}&= k_2^2k_3^2k_1k_4\Bigg[C_{14}C_{23}^{2} - C_{13}C_{23}\,C_{24} + C_{12}C_{24} + C_{13}C_{34}\Bigg],\\
Q^{\zeta}_{3 A'A'}&=  k_{1} k_{2} k_{3} k_{4} \Bigg[  C_{12}   C_{34} +C_{14} C_{23} \Bigg].
\end{align}
\subsection*{Electric Field}
In the first channel
\beq
\tilde{Q}^{\zeta}_{1AA}\= 4\, C_{12}\, C_{34},\\
\tilde{Q}^{\zeta}_{1AA'}\=2\, C_{12}\,\bigl(1 + C_{34}^{2}\bigr),\\
\tilde{Q}^{\zeta}_{1A'A}\= 2\, \bigl(1 + C_{12}^{2}\bigr)\, C_{34},\\
\tilde{Q}^{\zeta}_{1A'A'}\= (1 + C_{12}^{2})(1 + C_{34}^{2}).
\eeq
In the second channel
\beq
\tilde{Q}^{\zeta}_{1AA}\= C_{13} C_{24} + C_{12} C_{34},\\
\tilde{Q}^{\zeta}_{1AA'}\= C_{12} C_{23} - C_{23} C_{14} C_{24} + C_{13} C_{24}^{2} + C_{14} C_{34},\\
\tilde{Q}^{\zeta}_{1A'A}\= C_{12} C_{14} - C_{13} C_{23} C_{14} + C_{13}^{2} C_{24} + C_{23} C_{34},\\
\tilde{Q}^{\zeta}_{1A'A'}\=-1 + C_{12}^{2} + C_{13}^{2} + C_{23}^{2} + C_{14}^{2} + C_{24}^{2}+ C_{34}^{2}
- C_{34}\bigl(C_{13} C_{14} + C_{23} C_{24}\bigr)\\
&- C_{12}\,\l C_{14} C_{24} + C_{13}\,\l C_{23} - C_{24} C_{34}\r \r.
\eeq
In the third channel
\beq
\tilde{Q}^{\zeta}_{1AA}\= C_{23} C_{14} + C_{12} C_{34},\\
\tilde{Q}^{\zeta}_{1AA'}\= C_{23}^{2} C_{14} + C_{12} C_{24} - C_{13} C_{23} C_{24} + C_{13} C_{34},\\
\tilde{Q}^{\zeta}_{1A'A}\= C_{12} C_{13} + C_{23} C_{14}^{2} + C_{24}(-C_{13} C_{14} + C_{34}),\\
\tilde{Q}^{\zeta}_{1A'A'}\= -1 + C_{12}^{2} + C_{13}^{2} + C_{23}^{2} + C_{14}^{2} + C_{24}^{2} + C_{34}^{2}-  C_{34}\l C_{13} C_{14}+C_{23} C_{24}\r \\
&- C_{12}\,(C_{13} C_{23} + C_{14}(C_{24} - C_{23} C_{34})).
\eeq
\section{Integral results for different values of $n$}
\label{intresults}
In this appendix, we outline the method used to evaluate the nested time integrals appearing in the main text and present the results for different values of $n$ with $k_{1}=k_{2}=k$ and $k_{3}=k_{4}=p$. Using the series representation of the Hankel function, the evaluation of the nested time integral reduces to computing integral of the type
\beq
I_{n}\l a,\eta\r\= \int_{\eta}^{\infty}d\eta'\; \frac{e^{-ia \eta'}}{\eta'^{n}}.
\eeq
One can compute the above integral iteratively, knowing that
\beq
I_{n}\l a,\eta\r =
\begin{cases}
\Gamma\l 0,ia\eta\r, & \text{if } n = 1, \\
-\frac{i}{a}\, e^{-i a \eta}, & \text{if } n=0,\\
\frac{e^{-i a \eta}}{(n-1)\,\eta^{\,n-1}}
- \frac{i a}{n-1}\, I_{n-1}\l a, \eta\r & \text{if } n>1,\\
-\frac{i}{a}\, \eta^{-n} e^{-i a \eta}
+ \frac{i n}{a}\, I_{n+1}\l a, \eta\r & \text{if }  n<0,
\end{cases}
\eeq
where $\Gamma$ is the incomplete gamma function. In the computation one may also require to note that
\beq
\lim_{x\rightarrow 0}\Gamma\l 0,i x\r = -\gamma_{E}-\log\l ix\r.
\eeq
Using this we computed the nested time integral for different values of $n$ and is listed below.
In this section, we list the integral results used in the main text for different values of $n$ with $k_{1}=k_{2}=k$ and $k_{3}=k_{4}=p$.
\subsection{$n=0$}
\beq
I^{AA}\= |\zeta_{k_I}^{(0)}|^2|A_{k}^{(0)}|^{2}|A_{p}^{(0)}|^{2}\frac{1}{k p\left( 2k  + k_{\mathrm{I}} \right)^{2} }\bigg[\frac{\left(k +k_{\mathrm{I}} \right) \left(p + k_{\mathrm{I}} \right)}{\left( 2p + k_{\mathrm{I}} \right)^{2}}+ \frac{\l k+ k_{\mathrm{I}}\r}{k+p}\\
&-\frac{k_{\mathrm{I}}^{2}}{4\l k+p\r^{2}}-\frac{k_{\mathrm{I}}^{2}\l 2k+k_{\mathrm{I}}\r}{4\l k+p\r^{3}}\bigg] +  k \leftrightarrow p .
\eeq
\subsection{$n=1$}
The integral $I^{A'A'}$ for $n=1$ is obtained by multiplying $k^{2}p^{2}$, arising from the derivative, with $I^{AA}$ at $n=0$. Therefore, we provide only $I^{AA'}$ and $I^{AA}$ below, noting that $I^{A'A}$ follows from $I^{AA'}$ by exchanging $k$ and $p$.
\beq
I^{AA'}\= |\zeta_{k_I}^{(0)}|^2|A_{k}^{(0)}|^{2}|A_{p}^{(0)}|^{2} \frac{1}{4k p(k + p)^{3} (2k + k_{\mathrm{I}})^{2} (2p + k_{\mathrm{I}})^{2}}\big[ 8k_{\mathrm{I}}^{2} (k + p)^{3} (3 k^{2} + 4 k p + 3 p^{2})\\ 
&+24(k + p)^{3} (k^{2} + p^{2})\l kp + k_{\mathrm{I}}(k+p)\r  + 4 k_{\mathrm{I}}^{3}(5 k^{4} + 23 k^{3} p + 32 k^{2} p^{2} + 20 k p^{3} + 5 p^{4})\\
& + 4 k_{\mathrm{I}}^{4}  (k + p) (5 k^{2} + 6 k p + 2 p^{2})
+ k_{\mathrm{I}}^{5} (5 k^{2} + 6 k p + 2 p^{2}) \big],
\eeq
\beq
I^{AA} \= |\zeta_{k_I}^{(0)}|^2|A_{k}^{(0)}|^{2}|A_{p}^{(0)}|^{2}\frac{1}{8k^{3}p^{3}}\frac{k^{2}}{(k + p)^{3} (2k + k_{\mathrm{I}})^{2} (2p + k_{\mathrm{I}})^{2}}\big[4 k p k_{\mathrm{I}}^{3} \left( 135 p^{2} + 82 p k_{\mathrm{I}} + 9 k_{\mathrm{I}}^{2} \right) \\
&+23 p^{2} k_{\mathrm{I}}^{5} + 48 k^{4} \left( 6 p^{3} + 6 p^{2} k_{\mathrm{I}} + 4 p k_{\mathrm{I}}^{2} + k_{\mathrm{I}}^{3} \right)+ 48 k^{3} (3 p + k_{\mathrm{I}}) \left( 6 p^{3} + 6 p^{2} k_{\mathrm{I}} + 4 p k_{\mathrm{I}}^{2} + k_{\mathrm{I}}^{3} \right)\\
&+ 4 k^{2} k_{\mathrm{I}} \left( 216 p^{4} + 408 p^{3} k_{\mathrm{I}} + 208 p^{2} k_{\mathrm{I}}^{2} + 48 p k_{\mathrm{I}}^{3} + 3 k_{\mathrm{I}}^{4} \right)\big]+ k \leftrightarrow p .
\eeq
\subsection{$n=2$}
As in the $n=1$ case, we provide only $I^{AA}$ and $I^{AA'}$ below.
\beq
I^{A A'}\= |\zeta_{k_I}^{(0)}|^2|A_{k}^{(0)}|^{2}|A_{p}^{(0)}|^{2}\frac{1}{4k^{5}p\left(4 p^{2} - k_{\mathrm{I}}^{2} \right)^{2}}\ll 24 p^{3} k_{\mathrm{I}}^{3} \left(12 p^{2} - k_{\mathrm{I}}^{2} \right)\log \l \frac{2 (k + p)}{2 k + k_{\mathrm{I}}}\r\right.\\
&\left.- 16 k p^{3} \left( 9 k^{4} - 5 k^{3} k_{\mathrm{I}} + 18 k p^{2} k_{\mathrm{I}} - 36 p^{2} k_{\mathrm{I}}^{2} + 3 k_{\mathrm{I}}^{4} + k^{2} \left( -48 p^{2} + 4 k_{\mathrm{I}}^{2} \right) \right)\right.\\
&\left.-\frac{4 p^{2}k_{\mathrm{I}}^{3}}{(k + p)^{3}} \left( 15 k^{6} + 39 k^{5} p + 97 k^{4} p^{2} + 228 k^{3} p^{3} + 276 k^{2} p^{4} + 144 k p^{5} + 24 p^{6} \right)\right.\\
&\left. +\frac{ k_{\mathrm{I}}^{5}}{(k + p)^{3}}\left( 10 k^{6} + 30 k^{5} p + 65 k^{4} p^{2} + 122 k^{3} p^{3} + 150 k^{2} p^{4} + 96 k p^{5} + 24 p^{6} \right)\right.\\
&\left. + \frac{64}{(2 k + k_{\mathrm{I}})^{2}} \l \l 2k+k_{\mathrm{I}}\r\left( 4 k^{6} p^{3} + 15 k^{4} p^{5} \right)\r+ k^{5} p^{3} \left( k^{2} - 3 p^{2} \right) \rr ,
\eeq
\beq
I^{A A}\=|\zeta_{k_I}^{(0)}|^2|A_{k}^{(0)}|^{2}|A_{p}^{(0)}|^{2}\frac{k_{\mathrm{I}}^{6}}{4k^{5}p^{5}}\Bigg[6\pi^{2} +\frac{24 k \left( 20 k^{6} + 25 k^{4} k_{\mathrm{I}}^{2} - 20 k^{2} k_{\mathrm{I}}^{4} + 3 k_{\mathrm{I}}^{6} \right)}{k_{\mathrm{I}}^{3}\left( 2k - k_{\mathrm{I}} \right)^{2} \left( 2k + k_{\mathrm{I}} \right)^{2}}\log\!\left( \frac{2(k + p)}{2p + k_{\mathrm{I}}} \right)\\ &-18\operatorname{Li}_{2} \!\left(\frac{k_{\mathrm{I}}-2k}{2p+k_{\mathrm{I}}}\right)+\frac{1}{8k_{\mathrm{I}}^{6}(2p + k_{\mathrm{I}})^{2} (2k + k_{\mathrm{I}})^{2} (k + p)^{3}} 
\Big\{ 32\, k_{\mathrm{I}} (k + p)^{4}\left( -50 k^{4} p^{4} + 9 k_{\mathrm{I}}^{8} \right)\\
&+ 8k^{2} k_{\mathrm{I}}^{7}(k + p)^{2}\left( 317 k^{2} + 1280 kp + 960 p^{2} \right)-16 k^{5}p^{2} k_{\mathrm{I}}^{3}\left( 212 k^{3} + 796 k^{2} p + 1216 k p^{2} + 631 p^{3} \right)\\
&+ 16 k_{\mathrm{I}}^{4}k^{4}p (k + p)\left( 88 k^{3} + 216 k^{2} p - 104 k p^{2} - 231 p^{3} \right)-576(k + p)^{3} k k_{\mathrm{I}}^{8}(4 k + 3 p)\\
&  + 4k^{4}k_{\mathrm{I}}^{5}\left( 208 k^{4} + 1264 k^{3} p + 2932 k^{2} p^{2} + 3668 k p^{3} + 1793 p^{4} \right)+1600 k^{5} p^{5}(k + p)^{3}\\
&+ 8(k + p)^{3} k \Big(200 k^{3} p^{3}k_{\mathrm{I}}^{2} (4 k + p) -k \left( 110 k^{2} + 902 k p + 381 p^{2} \right) k_{\mathrm{I}}^{6}  
\Big)\Big\}\Bigg]+  k \leftrightarrow p .
\eeq
When all $k'$s are equal, i.e.,  $k_{1}=k_{2}=k_{3}=k_{4}=k$ and  $x = k_{\mathrm{I}}/k$
\beq
I^{A' A'} \= |\zeta_{k_I}^{(0)}|^2|A_{k}^{(0)}|^{2}|A_{p}^{(0)}|^{2} \frac{1}{32 (2 + x)^{4}}
\l 1152 + 2304 x + 2688 x^{2} + 1756 x^{3} + 568 x^{4} + 71 x^{5}\r ,
\eeq
\beq
I^{A A'} \= |\zeta_{k_I}^{(0)}|^2|A_{k}^{(0)}|^{2}|A_{p}^{(0)}|^{2}\frac{9\, x^{3}}{4 k^{2}}\ll \frac{1}{72 x^{3}\left(2- x\right) \left(2 + x\right)^{4}}\l 3840  + 5760 x + 10240 x^{2} + 5896 x^{3}\right.\right.\\
&\left.\left.  - 2356 x^{4} - 2598 x^{5} - 497 x^{6}\r +\frac{8\l 12-x^{2}\r}{3 \left(4 - x^{2}\right)^{2}}\log\l\frac{4}{2+x}\r \rr ,
\eeq
\beq 
I^{A A}&= |\zeta_{k_I}^{(0)}|^2|A_{k}^{(0)}|^{2}|A_{p}^{(0)}|^{2}\frac{9x^{6}}{k^{4}}\Big[ \frac{4\left(20 + 25 x^{2} - 20 x^{4} + 3 x^{6}\right)}{3\,x^{3}\left(4- x^{2}\right)^{2}}\log\left(\frac{4}{2 + x}\right)-\operatorname{Li}_{2}\left(\frac{2 + x}{x - 2}\right)\\ 
& \frac{1}{72 x^{6} \left(2 - x\right)\left(2 + x\right)^{4}}\left( -70080 - 66904 x + 3308 x^{2} + 18258 x^{3} + 4323 x^{4} \right) \\ 
&\frac{1}{288 x^{6}} \left(8960 - 4777 x + 3072 x^{2} - 2048 x^{3} + 1152 x^{4} - 1152 x^{5} + 48 \pi^{2} x^{6}\right)\Big].
\eeq
\subsection{$n=3$}
Here we focus on the dominant contributions of the integral in the $\eta_{0}\to 0$ limit. Thus, we present only $I^{AA}$ in the equilateral case, i.e., $k_{1}=k_{2}=k_{3}=k_{4}=k$. 
\beq
I^{A A} \=  |\zeta_{k_I}^{(0)}|^2|A_{k}^{(0)}|^{2}|A_{p}^{(0)}|^{2}\frac{225}{4 k^{14}}\ll \frac{4kk_{\mathrm{I}}^{3}\log\l\frac{4k}{2k+k_{\mathrm{I}}}\r}{15\l 2k-k_{\mathrm{I}}\r^{2}\l 2k+k_{\mathrm{I}}\r^{4}}\left( 32 k^{12} - 352 k^{11} k_{\mathrm{I}} - 608 k^{10} k_{\mathrm{I}}^{2}\right.\right.\\
&\left.\left.- 48 k^{9} k_{\mathrm{I}}^{3} + 378 k^{8} k_{\mathrm{I}}^{4} + 182 k^{7} k_{\mathrm{I}}^{5}
 - 29 k^{6} k_{\mathrm{I}}^{6} - 2352 k^{5} k_{\mathrm{I}}^{7} + 52 k^{4} k_{\mathrm{I}}^{8} + 640 k^{3} k_{\mathrm{I}}^{9} + 100 k^{2} k_{\mathrm{I}}^{10}\right.\right.\\
 &\left.\left.- 60 k k_{\mathrm{I}}^{11} - 15 k_{\mathrm{I}}^{12} \right)-k_{\mathrm{I}}^{6} \left(-2 k^{2} + k_{\mathrm{I}}^{2}\right)^{2}
\operatorname{Li}_{2}\!\left( \frac{k_{\mathrm{I}}-2k}{2k + k_{\mathrm{I}}}\right)+\frac{1}{3} \pi^{2} q^{6} \bigl(-2 k^{2} + k_{\mathrm{I}}^{2}\bigr)^{2}\right.\\
&\left.-\frac{8}{35}\,\gamma_{\mathrm{E}}\, k^{7} k_{\mathrm{I}}^{3}-\frac{15136}{225} k^{10} + \frac{7049}{200} k^{9} k_{\mathrm{I}} - \frac{176}{9} k^{8} k_{\mathrm{I}}^{2} + \frac{120544}{11025} k^{7} k_{\mathrm{I}}^{3}- \frac{272}{45} k^{6} k_{\mathrm{I}}^{4}\right.\\
&\left.+ \frac{1616}{225} k^{5} k_{\mathrm{I}}^{5} + \frac{16}{3} k^{4} k_{\mathrm{I}}^{6} - \frac{80}{9} k^{3} k_{\mathrm{I}}^{7} - 4 k^{2} k_{\mathrm{I}}^{8} + 4 k k_{\mathrm{I}}^{9} - \frac{8}{225} \frac{k^{14}}{(2k+k_{\mathrm{I}})^{4}} - \frac{176}{225} \frac{k^{13}}{(2k+k_{\mathrm{I}})^{3}}\right.\\
&\left. + \frac{463}{45} \frac{k^{12}}{(2k+k_{\mathrm{I}})^{2}} + \frac{k^{11}}{450} \left(\frac{480}{-2k+k_{\mathrm{I}}} + \frac{58603}{2k+k_{\mathrm{I}}}\right) - \frac{4\, k^{7} k_{\mathrm{I}}^{3}\log\!\left(-4 k \eta_{0}\right)}{105 \left(-2 k + k_{\mathrm{I}}\right)^{2} \left(2 k + k_{\mathrm{I}}\right)^{4}}\left(2976 k^{6}\right.\right.\\
&\left.\left.+ 288 k^{5} k_{\mathrm{I}} + 5248 k^{4} k_{\mathrm{I}}^{2} + 8480 k^{3} k_{\mathrm{I}}^{3} - 18638 k^{2} k_{\mathrm{I}}^{4} - 22214 k k_{\mathrm{I}}^{5} + 10389 k_{\mathrm{I}}^{6}\right)\right.\\
&\left.+ \frac{240\, k^{7} k_{\mathrm{I}}^{3}\log\!\big(-\bigl(2k + k_{\mathrm{I}}\bigr)\eta_{0}\big)}{\left(-2k + k_{\mathrm{I}}\right)^{2} \left(2k + k_{\mathrm{I}}\right)^{4}} \big(480 k^{6} + 96 k^{5} k_{\mathrm{I}} + 736 k^{4} k_{\mathrm{I}}^{2} + 1184 k^{3} k_{\mathrm{I}}^{3} - 2666 k^{2} k_{\mathrm{I}}^{4}\right.\\
&\left.- 3170 k k_{\mathrm{I}}^{5} + 1485 k_{\mathrm{I}}^{6}\big)\rr .
\eeq
Note that this integral behaves as $\log\l \eta_{0}\r$ in the $\eta_{0}\rightarrow 0$ limit due to the last two terms. To make it explicit, one can write this as
\beq
I^{AA}=|\zeta_{k_I}^{(0)}|^2|A_{k}^{(0)}|^{2}|A_{p}^{(0)}|^{2}\ll I^{AA}_{\text{finite}}\l k,k_{\mathrm{I}}\r +\frac{90 k_{\mathrm{I}}^{3}}{7k^{7}}\log\l -\eta_{0}\r \rr .
\eeq
\section{Integral identity evaluation using equations of motion}
\label{intidentity}
We begin by defining the integral of interest as
\beq
I^{1XY} \equiv 2\,\text{Re}\!\left[I^{1XY}_{++}+I^{1XY}_{-+}\right].
\eeq
This form captures the physical contribution from both the ``$++$'' and ``$-+$'' branches of the in-in contour, with the overall factor of two ensuring that we extract the real part of the correlation function.  
In what follows, we first focus on $I^{XY}_{++}$ (now onward, let’s suppress the channel superscript $1$), and then extend the analysis to $I^{XY}_{-+}$ to obtain the complete result.
In the counter collinear limit, $k_{\mathrm{I}}\to 0$, the gravitational mode functions simplify considerably as they become time-independent.  
This property allows us to reduce the nested double time integral into a product of two single time integrals involving only the gauge field mode functions.  
Explicitly, we obtain
\beq
I^{A'A'}_{++} = -|\zeta_{k_{\mathrm{I}}}|^2A_{k}^{2}(\eta_{0})\,A_{p}^{2}(\eta_{0})
   \int_{-\infty}^{\eta_{0}} d\eta_{1}\,\lambda(\eta_{1})\,A_{k}^{*'}(\eta_{1})A_{k}^{*'}(\eta_{1})
   \int_{-\infty}^{\eta_{0}} d\eta_{2}\,\lambda(\eta_{2})\,A_{p}^{*'}(\eta_{2})A_{p}^{*'}(\eta_{2})~.
\eeq
Here, the prefactors $A_{k}(\eta_{0})$ and $A_{p}(\eta_{0})$ arise from the external legs of the diagram, while the integrals over $\eta_{1}$ and $\eta_{2}$ encode the bulk interaction history of the gauge fields.
Our next step is to re-express $I^{A'A'}_{++}$ in terms of the simpler integral $I^{AA}_{++}$.  
To this end, we exploit the equation of motion for the gauge field mode function,
\beq
\frac{d}{d\eta}\!\left(\lambda\,A_{k}'\right) = -k^{2}\lambda\,A_{k},
\eeq
which encapsulates the dynamics of the system in the presence of the time-dependent coupling $\lambda(\eta)$.  
This relation allows us to trade derivatives of the mode functions for algebraic factors of the momentum $k$, which will be crucial in simplifying the structure of the integrals.
Applying integration by parts, we find
\begin{align}
\int_{-\infty}^{\eta_{0}} d\eta_{1}\,\lambda(\eta_{1})\,A_{k}^{*'}(\eta_{1})A_{k}^{*'}(\eta_{1})
&= \lambda(\eta_{0})\,A^{*'}_{k}(\eta_{0})A^{*}_{k}(\eta_{0})
   + k^{2}\!\int_{-\infty}^{\eta_{0}} d\eta\,\lambda(\eta)\,A_{k}^{*}(\eta)A_{k}^{*}(\eta) \nonumber \\
&\equiv \lambda \,A^{*'}_{k}A^{*}_{k}
   + k^{2} I^{0}_{k},
\end{align}
where we have introduced $I^{0}_{k}$ for notational convenience.  
The first term represents a boundary contribution evaluated at the upper limit $\eta_{0}$ and lower limit vanishes in the prescribed time contour. From now on, we suppress this time arguments since it is trivial. The second term corresponds to a bulk integral that is structurally similar to $I^{AA}_{++}$.  
Substituting this back into the expression for $I^{A'A'}_{++}$, we obtain
\begin{align}
I^{A'A'}_{++} &= -|\zeta_{k_{\mathrm{I}}}|^2\,A_{k}^{2}\,A_{p}^{2}
   \Bigg[ \big(\lambda \,A^{*'}_{k} A^{*}_{k} + k^{2}I^{0}_{k}\big)
          \big(\lambda\,A^{*'}_{p}A^{*}_{p} + p^{2}I^{0}_{p}\big)\Bigg].
\end{align}
Expanding this product, we can identify contributions equal to $I^{AA}_{++}$, as well as mixed terms involving boundary and bulk pieces:
\begin{align}
I^{A'A'}_{++}
&= k^{2}p^{2}I^{AA}_{++}-|\zeta_{k_{\mathrm{I}}}|^2\,A_{k}^{2}\,A_{p}^{2}
   \Bigg[ \lambda^{2}\,A^{*'}_{k}A^{*}_{k}A^{*'}_{p}A^{*}_{p} + p^{2}\lambda \,A^{*'}_{k}A^{*}_{k}I^0_{p}
          + k^{2}\lambda\,A^{*'}_{p}A^{*}_{p}I^0_{k} \Bigg].
\end{align}
This decomposition makes explicit how $I^{A'A'}_{++}$ is built from a combination of bulk integrals ($I^{AA}_{++}$ and $I^{0}_{k,p}$) and boundary contributions evaluated at $\eta_{0}$.
Proceeding in the same manner, we obtain for the mixed integrals
\begin{align}
I^{AA'}_{++} &=p^{2}I^{AA}_{++} -|\zeta_{k_{\mathrm{I}}}|^2\,A_{k}^{2}\,A_{p}^{2}\lambda\,A^{*'}_{p}A^{*}_{p}I^0_{k},\\
I^{AA'}_{++} &=k^{2}I^{AA}_{++} -|\zeta_{k_{\mathrm{I}}}|^2\,A_{k}^{2}\,A_{p}^{2}
\lambda\,A^{*'}_{k}A^{*}_{k}I^0_{p}.
\end{align}
These relations again demonstrate the interplay between bulk and boundary terms in the evaluation of the nested time integrals.
Collecting the results, we find the following compact identity
\beq
I^{A'A'}_{++} + k^{2}p^{2}I^{AA}_{++} - k^{2}I^{AA'}_{++} - p^{2}I^{A'A}_{++}
= -\lambda^{2}\,|\zeta_{k_{\mathrm{I}}}|^2|A_{k}|^{2}|A_{p}|^{2}\,A_{k}A^{*'}_{k}A_{p}A^{*'}_{p}.
\eeq
Physically, this relation shows that a precise linear combination of the various integrals collapses into a purely boundary-localized expression, reflecting the underlying constraints imposed by the equations of motion.
By repeating the same steps for the ``$-+$'' branch, we obtain
\beq
I^{A'A'}_{-+}+k^{2}p^{2}I^{AA}_{-+}-k^{2}I^{AA'}_{-+}-p^{2}I^{A'A}_{-+}
= \lambda^{2}\,|\zeta_{k_{\mathrm{I}}}|^2|A_{k}|^{2}|A_{p}|^{2}\,A^{*}_{k}A_{k}'A_{p}A^{*'}_{p}.
\eeq
The structure is similar to the $++$ case, but differs by a relative sign and complex conjugation due to the in-in contour prescription. Finally, combining both contributions, we arrive at the expression for the four-point autocorrelation function of the magnetic fields in the counter collinear limit.
\bibliography{references}
\bibliographystyle{JHEP}
\end{document}